\documentclass[twocolumn]{aastex63}
\usepackage[utf8]{inputenc}
\usepackage{hyperref}

\usepackage{float}
\usepackage{verbatim}
\usepackage{natbib}

\accepted{Astrophysical Journal}
\usepackage{xcolor}
\newcommand\ticstar{TIC\,454140642}

\begin{document}

\title{TIC 454140642: A Compact, Coplanar, Quadruple-lined Quadruple Star System Consisting of Two Eclipsing Binaries}

\correspondingauthor{Veselin Kostov}
\email{veselin.b.kostov@nasa.gov}
%
\author[0000-0001-9786-1031]{Veselin~B.~Kostov}
\affiliation{NASA Goddard Space Flight Center, 8800 Greenbelt Road, Greenbelt, MD 20771, USA}
\affiliation{SETI Institute, 189 Bernardo Ave, Suite 200, Mountain View, CA 94043, USA}
\affiliation{GSFC Sellers Exoplanet Environments Collaboration}
%
\author[0000-0003-0501-2636]{Brian P. Powell}
\affiliation{NASA Goddard Space Flight Center, 8800 Greenbelt Road, Greenbelt, MD 20771, USA}
%
\author[0000-0002-5286-0251]{Guillermo Torres}
\affiliation{Center for Astrophysics $\vert$ Harvard \& Smithsonian, 60 Garden St, Cambridge, MA, 02138, USA}
%
\author[0000-0002-8806-496X]{Tam\'as Borkovits}
\affiliation{Baja Astronomical Observatory of University of Szeged, H-6500 Baja, Szegedi út, Kt. 766, Hungary}
\affiliation{Konkoly Observatory, Research Centre for Astronomy and Earth Sciences, H-1121 Budapest, Konkoly Thege Miklós út 15-17, Hungary}
\affiliation{ELTE Gothard Astrophysical Observatory, H-9700 Szombathely, Szent Imre h. u. 112, Hungary}
%
\author[0000-0003-3182-5569]{Saul A. Rappaport}
\affiliation{Department of Physics, Kavli Institute for Astrophysics and Space Research, M.I.T., Cambridge, MA 02139, USA}
\author[0000-0002-2084-0782]{Andrei Tokovinin}
\affiliation{Cerro Tololo Inter-American Observatory | NSF's NOIRab, Casilla 603, La Serena, Chile}
%
\author[0000-0001-9383-7704]{Petr Zasche}
\affiliation{Astronomical Institute, Charles University, Faculty of Mathematics and Physics, V Hole\v{s}ovi\v{c}k\'ach 2, CZ-180 00, Praha 8, Czech Republic}
%
%
\author{David Anderson}
\affiliation{Department of Physics, University of Warwick, Gibbet Hill Road, Coventry CV4 7AL, UK}
%
\author[0000-0001-7139-2724]{Thomas~Barclay}
\affiliation{NASA Goddard Space Flight Center, 8800 Greenbelt Road, Greenbelt, MD 20771, USA}
\affiliation{University of Maryland, Baltimore County, 1000 Hilltop Cir,
Baltimore, MD 21250, USA}
\author{Perry Berlind}
\affiliation{Center for Astrophysics $\vert$ Harvard \& Smithsonian, 60 Garden St, Cambridge, MA, 02138, USA}
%
\author[0000-0002-3481-9052]{Peyton Brown}
\affiliation{Department of Physics and Astronomy, Vanderbilt University, 6301 Stevenson Center Ln., Nashville, TN 37235, USA}
\author{Michael L.\ Calkins}
\affiliation{Center for Astrophysics $\vert$ Harvard \& Smithsonian, 60 Garden St, Cambridge, MA, 02138, USA}
%
\author[0000-0001-6588-9574]{Karen A.\ Collins}
\affiliation{Center for Astrophysics $\vert$ Harvard \& Smithsonian, 60 Garden St, Cambridge, MA, 02138, USA}
%
\author[0000-0003-2781-3207]{Kevin I.\ Collins}
\affiliation{George Mason University, 4400 University Drive, Fairfax, VA, 22030 USA}
%
\author[0000-0003-2239-0567]{Dennis M.\ Conti}
\affiliation{American Association of Variable Star Observers, 49 Bay State Road, Cambridge, MA 02138, USA}
\author{Gilbert A. Esquerdo}
\affiliation{Center for Astrophysics $\vert$ Harvard \& Smithsonian, 60 Garden St, Cambridge, MA, 02138, USA}
%
\author{Coel Hellier}
\affil{Astrophysics Group, Keele University, Staffordshire, ST5 5BG, UK}
%
\author[0000-0002-4625-7333]{Eric L.\ N.\ Jensen}
\affiliation{Department of Physics \& Astronomy, Swarthmore College, Swarthmore PA 19081, USA}
%
\author{Jacob Kamler}
\affiliation{John F. Kennedy High School, 3000 Bellmore Avenue, Bellmore, NY 11710, USA}
%
\author[0000-0002-0493-1342]{Ethan Kruse}
\affiliation{NASA Goddard Space Flight Center, 8800 Greenbelt Road, Greenbelt, MD 20771, USA}
\affiliation{Universities Space Research Association, 7178 Columbia Gateway Drive, Columbia, MD 21046}
%
\author[0000-0001-9911-7388]{David W. Latham}
\affiliation{Center for Astrophysics $\vert$ Harvard \& Smithsonian, 60 Garden St, Cambridge, MA, 02138, USA}
\author[0000-0002-0967-0006]{Martin Ma\v{s}ek}
\affiliation{FZU - Institute of Physics of the Czech Academy of Sciences, Na Slovance 1999/2, CZ-182 21, Praha, Czech Republic}
%
\author{Felipe Murgas}
\affiliation{Instituto de Astrof\'isica de Canarias (IAC), E-38205 La Laguna, Tenerife, Spain}
\affiliation{Departamento de Astrof\'isica, Universidad de La Laguna (ULL), E-38206 La Laguna, Tenerife, Spain}
%
\author[0000-0001-8472-2219]{Greg Olmschenk}
\affiliation{NASA Goddard Space Flight Center, 8800 Greenbelt Road, Greenbelt, MD 20771, USA}
\affiliation{Universities Space Research Association, 7178 Columbia Gateway Drive, Columbia, MD 21046}
%
\author[0000-0001-9647-2886]{Jerome A. Orosz}
\affil{Department of Astronomy, San Diego State University, 5500 Campanile Drive, San Diego, CA 92182, USA}
%
\author[0000-0001-5449-2467]{Andr\'as P\'al}
\affiliation{Konkoly Observatory, Research Centre for Astronomy and Earth Sciences, MTA Centre of Excellence, Konkoly Thege Mikl\'os  \'ut 15-17, H-1121 Budapest, Hungary}
%
\author{Enric Palle}
\affiliation{Instituto de Astrof\'isica de Canarias (IAC), E-38205 La Laguna, Tenerife, Spain}
\affiliation{Departamento de Astrof\'isica, Universidad de La Laguna (ULL), E-38206 La Laguna, Tenerife, Spain}
%
\author[0000-0001-8227-1020]{Richard P. Schwarz}
\affiliation{Patashnick Voorheesville Observatory, Voorheesville, NY 12186, USA}
%
\author[0000-0003-2163-1437]{Chris Stockdale}
\affiliation{Hazelwood Observatory, Australia}
%
\author{Daniel Tamayo}
\affiliation{Department of Astrophyiscal Sciences, Princeton University, Princeton, NJ, 08544}
\affiliation{NASA Sagan Postdoctoral Fellow}
\affiliation{Lyman Spitzer Jr. Fellow}
\author{Robert Uhla\v{r}}
\affiliation{Private Observatory, Poho\v{r}\'{\i} 71, CZ-254 01 J\'{\i}lov\'e u Prahy, Czech Republic}
%
\author[0000-0003-2381-5301]{William F. Welsh}
\affil{Department of Astronomy, San Diego State University, 5500 Campanile Drive, San Diego, CA 92182, USA}
\author{Richard West}
\affiliation{Department of Physics, University of Warwick, Gibbet Hill Road, Coventry CV4 7AL, UK}
%
%
%
%

\begin{abstract}
We report the discovery of a compact, coplanar, quadruply-lined, eclipsing quadruple star system from TESS data, \ticstar, also known as TYC 0074-01254-1. The target was first detected in Sector 5 with 30-min cadence in Full-Frame Images and then observed in Sector 32 with 2-min cadence.  The light curve exhibits two sets of primary and secondary eclipses with periods of $\rm P_A = 13.624$ days (binary A) and $\rm P_B = 10.393$ days (binary B). Analysis of archival and follow-up data shows clear eclipse-timing variations and divergent radial velocities, indicating dynamical interactions between the two binaries and confirming that they form a gravitationally-bound quadruple system with a 2+2 hierarchy. The Aa+Ab binary, Ba+Bb binary, and A-B system are aligned with respect to each other within a fraction of a degree: the respective mutual orbital inclinations are 0.25 degrees (A vs B), 0.37 degrees (A vs A-B), and 0.47 degrees (B vs A-B). The A-B system has an orbital period of 432 days -- the second shortest amongst confirmed quadruple systems -- and an orbital eccentricity of 0.3.  
\end{abstract}


\section{Introduction}\label{sec:intro}

A few percent of F- and G-type stars are members of stellar quadruples, and the fraction likely increases with stellar mass \citep[e.g][]{Raghavan2010,DeRosa2014,Toonen2016,Moe2017,Tokovinin2017}. These systems are ideal laboratories to study stellar formation and evolution in a dynamically-complex environment where the constituent binary stars can be close enough to interact with each other. These interactions can occur multiple times over the lifetime of the system, involve e.g. Lidov-Kozai oscillations (Lidov 1962; Kozai 1962; Pejcha et al. 2013), mass-transfer, tidal and dynamical interactions, mergers (Perets \& Fabrycky 2009, Hammers et al. 2021), or formation of tight binaries (e.g. Naoz \& Fabrycky 2014, and references therein). Quadruple systems are important participants in stellar evolution as a potential source for Type I SNe (Fang et al. 2018), as well as interesting outcomes such as common-envelope evolution, interactions with circumbinary disks, etc.

Detection and detailed characterization of a bona-fide quadruple-star system can be challenging. For example, it is not uncommon for close eclipsing binary stars to exhibit deviations from strict periodicity and recent studies show that at least $\sim7\%$ of Kepler's eclipsing binary stars show such deviations within the 4-year-long observing window of the Kepler prime mission (e.g. Rappaport et al. 2013, Orosz 2015, Borkovits et al. 2016). Such eclipse timing variations (ETVs) can be indicative of dynamical interactions between the eclipsing binary and additional star(s) on wider orbit(s), and provide an important detection method for triple, quadruple, and higher-order stellar systems, as well as for circumbinary planets (e.g. Borkovits et al. 2016, Welsh \& Orosz 2018). Other pathways for the detection of quadruple (and higher-order multiples) are visual and spectroscopic observations (e.g. Raghavan et al. 2010, Tokovinin 2017, Pribulla et al. 2006), but these methods generally do not guarantee that the potentially additional star(s) are indeed gravitationally-bound to the known binary. 

A target star exhibiting multiple sets of eclipses, ETVs, and per-EB divergent radial velocities provides a unique opportunity to not only confirm the multiplicity of the system but also measure the orbital and physical parameters of its constituents with exquisite precision (e.g.~\citealt{Carter2011,Borkovits2018,Borkovits2020,Borkovits2021}). We note that because the outer orbital period of quadruple systems is much longer than the periods of the inner binaries (for otherwise the system would become unstable), detecting such features requires continuous and long-baseline observations of the target. A combination of space-based photometric surveys, extensive ground-based surveys photometry, and dedicated spectroscopy follow-up is ideally-suited to provide such observations. This was demonstrated by the highly-successful Kepler mission which detected a number of EB systems exhibiting additional features indicative of either circumbinary planets or triple and higher-order stellar systems (e.g. Welsh \& Orosz 2018; Kirk et al. 2015, Borkovits et al. 2016). 

\begin{figure}
    \centering
    \includegraphics[width=1.0\linewidth]{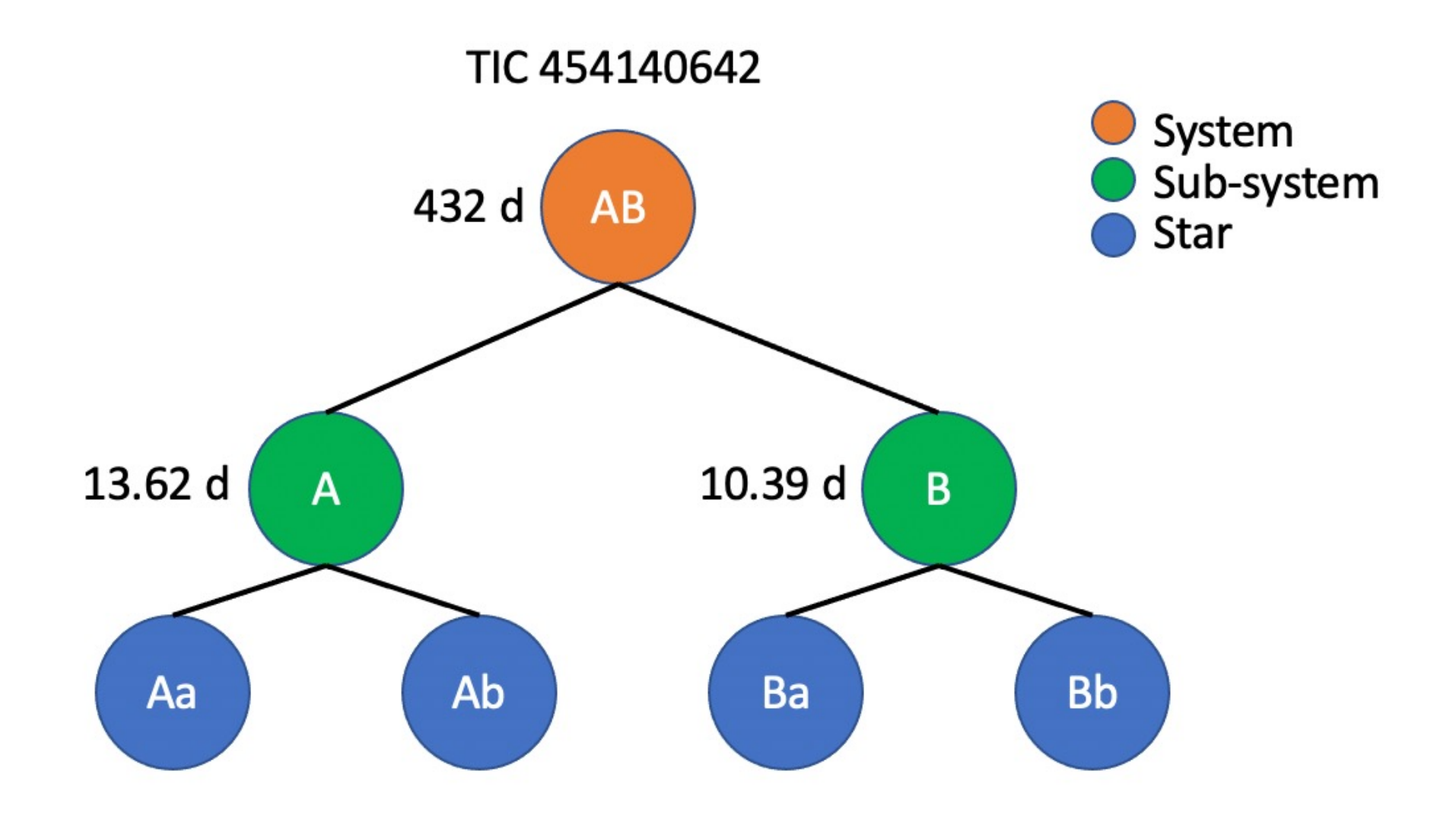}
    \caption{Structure of TIC 454140642, a quadruple star system consisting of two eclipsing binaries.}
   \label{fig:structure}
\end{figure}

The Transiting Exoplanet Survey Satellite (TESS, Ricker et al. 2015) is equally capable of discovering multiply-eclipsing stellar systems, as already demonstrated by \citet{Borkovits2020b,Borkovits2021} for stellar triples and quadruples, and \citet{Powell2021} for sextuples. The target presented here, \ticstar, exhibits the typical 2+2 hierarchy for quadruple systems, illustrated in Fig.~\ref{fig:structure}, and joins the small number of such confirmed, well-characterized systems --- VW LMi \citep{Pribulla2008}, V994 Her \citep{Zasche2016}, V482 Per (Torres et al. 2017), EPIC 220204960 \citep{Rappaport2017}, EPIC 219217635 \citep{Borkovits2018}, CzeV1731 \citep{zasche2020}, TIC 278956474 \citep{Rowden2020}, and BG Ind \citep{Borkovits2021}. 

We note that while quadruple star systems have lower occurrence rates compared to triple systems, the larger outer orbit necessary for stability in the latter systems is generally well beyond the 27-day TESS observation time in a single sector. Specifically, the long outer orbital period substantially reduces the probability that tertiary eclipses or occultations of a stellar triple will be identified in one sector of TESS data. While such triples can (and do) produce multiple tertiary eclipses or occultations during one conjunction of the outer orbit in a singe sector of TESS data \citep[e.g.][]{Borkovits2020b}, detecting multiple such conjunctions in one sector of data is highly unlikely. In turn, this makes determining the outer period of such triples -- and thus the overall configuration of the respective system -- challenging. In contrast, quadruple systems do not need to be at a conjunction of the outer orbit to reveal their nature, and eclipses/occultations of their constituent binaries can (and do) easily occur in a single sector of TESS data. 

Here we present the discovery of the quadruple star \ticstar\ that exhibits four sets of eclipses as well as prominent, dynamically induced apsidal motion, and short-term four-body perturbations. At the time of writing, \ticstar\ is the closest to coplanarity amongst the quadruple systems mentioned above, and has the second shortest outer period (after VW LMi). This paper is organized as follows. Section \ref{sec:detection} outlines the detection of the target in TESS data. In Sections \ref{sec:archival} and \ref{sec:follow-up} we describe the analysis of the available archival data and our follow-up observations, respectively. In Section \ref{sec:sysparms} we present our comprehensive spectro-photodynamical analysis of the system's parameters. We discuss the properties of the system in Section \ref{sec:discussion} and draw our conclusions in Section \ref{sec:summary}.  


\section{Detection}
\label{sec:detection}

Our detection methodology is described in detail by \citet{Powell2021}. Briefly, we found hundreds of thousands of eclipsing binaries using a neural network classifier on the TESS Full-Frame Image (FFI) light curves.  Through manual examination of the lightcurves identified by the neural network as EBs, we found many that showed eclipses belonging to multiple periods in the same light curve. Through photocenter analysis, we determined that the large majority of these lightcurves are consistent with a superposition of two EBs originating from unrelated target stars within the same TESS aperture. A fraction of these, however, demonstrated on-target photocenter for all detected eclipses, suggesting they originate from the same stellar system.  

\begin{figure*}
    \centering
    \includegraphics[width=1.0\linewidth]{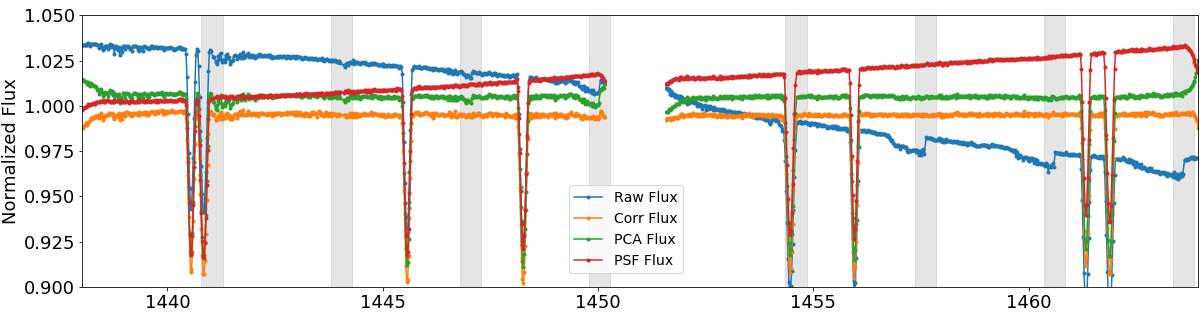}
    \includegraphics[width=1.0\linewidth]{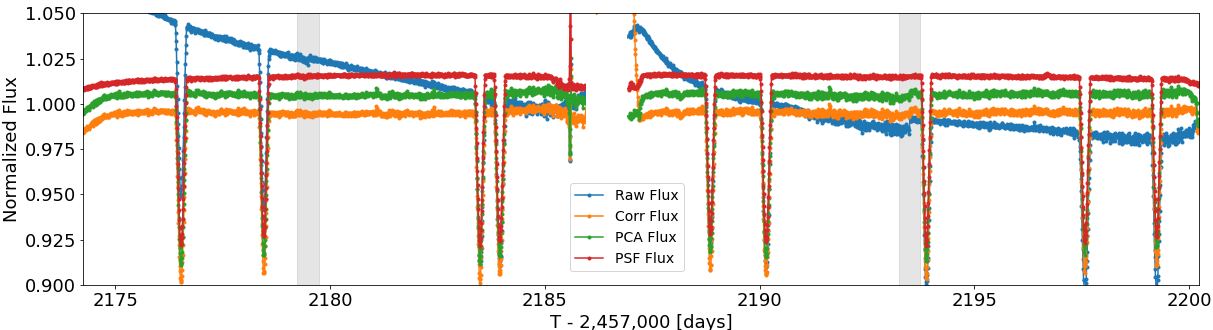}
    \caption{TESS photometry of \ticstar, constructed using {\tt eleanor} \citep{eleanor}, showing the raw flux (blue), corrected flux (orange), PCA flux (green), and PSF flux (red) lightcurve. The eclipses are present in all four lightcurves and are not strongly affected by known systematics such as momentum dumps (vertical grey bands) or background artifacts (spike near 2185.5). The upper panel shows Sector 5 data, the lower panel shows Sector 32 data. There are no apparent out-of-eclipse variations.}
   \label{fig:tess_lc}
\end{figure*}

\begin{figure*}
    \centering
    \includegraphics[width=0.295\linewidth]{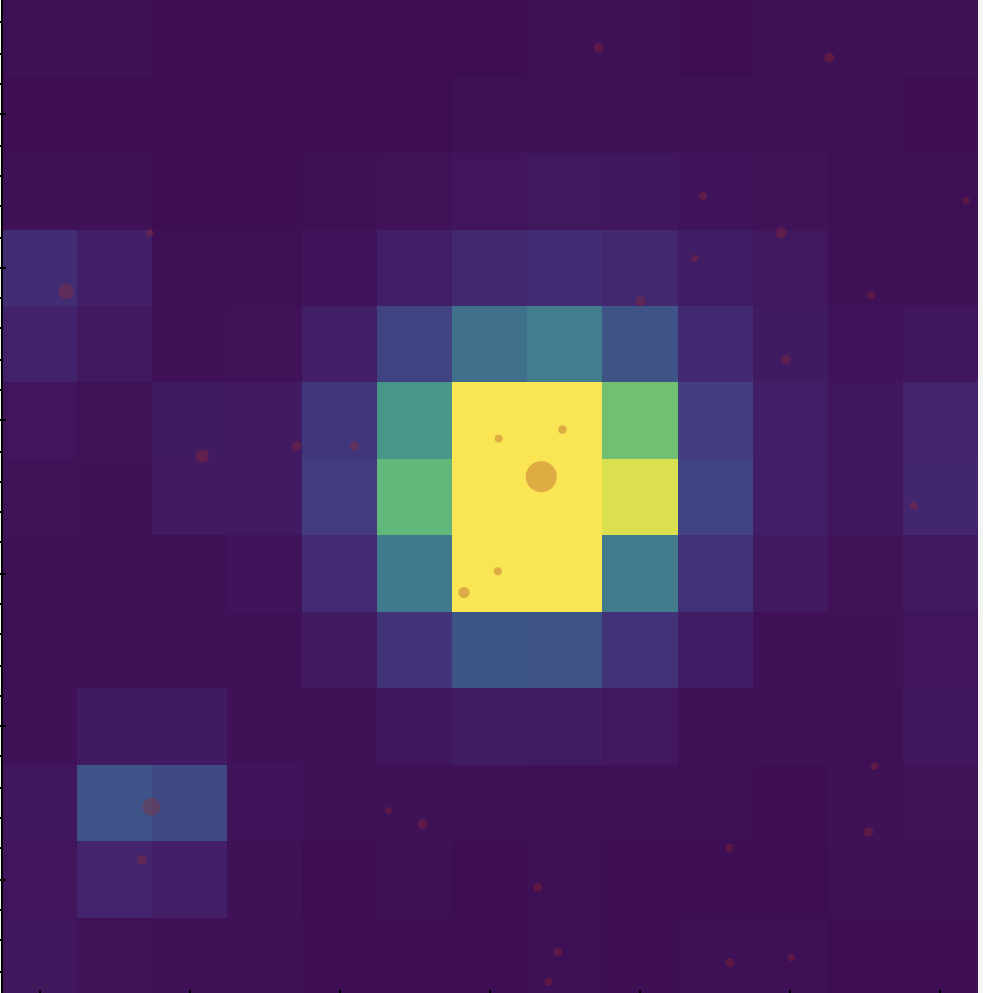} \hglue0.1cm
    \includegraphics[width=0.3\linewidth]{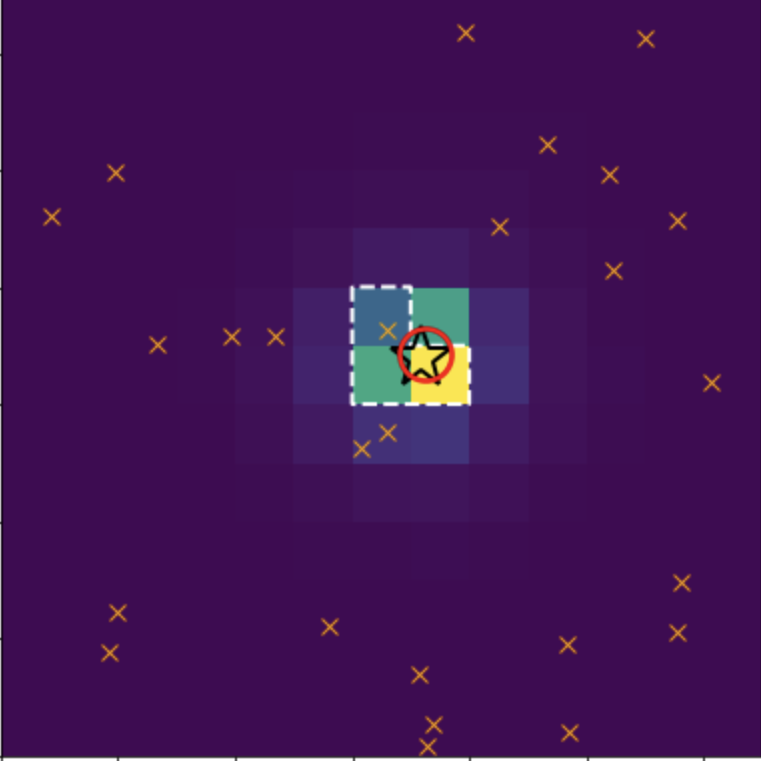} \hglue0.1cm
    \includegraphics[width=0.3\linewidth]{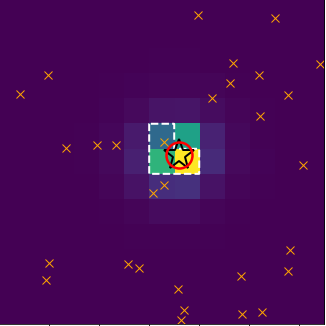} \hglue0.1cm
    \caption{Left: TESS aperture image of \ticstar\ ($T = 9.855$ mag) overlaid with Skyview sources brighter than $T = 21$ mag. Center: Mean difference image (out-of-eclipse minus in-eclipse) showing the measured average center-of-light of binary A eclipses (large red symbol), and the catalog position of the target (black star). Right: same as the left panel but for the binary B eclipses. All four sets of eclipses are on-target. The orange crosses represent nearby stars within ${\rm \Delta T= 10~mag}$. }
    \label{fig:photocenter}
\end{figure*}

\ticstar, a previously unknown EB, was one of the systems that we found early in our examination of TESS light curves. The nature of the system was initially quite challenging to decipher because all the eclipses are similar in depth and duration, in addition to the fact we only had one sector of data with each eclipse present twice in the lightcurve. Preliminary analysis indicated that there are two sets of primary and secondary eclipses. The TESS {\tt eleanor} lightcurve \citep{eleanor} for Sectors 5 and 32 is shown in Fig. \ref{fig:tess_lc}, highlighting the four sets of eclipses; there are no visible out-of-eclipse variations. In accordance with the designation system developed by the IAU, these binaries will be referred to as A (the brighter one with a period of 13.62 days, and components Aa and Ab) and B (the fainter one, with a period of 10.39 days and components Ba and Bb). The contamination due to nearby sources is low according to the TESS Input Catalog (0.001), and photocenter analysis confirmed that both sets of eclipses originate from the target (see Fig. \ref{fig:photocenter}), indicating a potential quadruple system. This motivated us to submit the target for 2-min cadence observations in Sector 32 (TESS DDT program 020 PI: Kostov), and also coordinate follow-up efforts with the TESS Follow-up Observing Program (TFOP). The parameters of the system are listed in Table \ref{tab:EBparameters}. 

\begin{deluxetable}{l r r r }[!ht]
\tabletypesize{\scriptsize}
\tablecaption{Stellar parameters for TIC 454140642\label{tab:EBparameters}}
\tablewidth{0pt}
\tablehead{
\colhead{Parameter} & \colhead{Value} & \colhead{Error} &\colhead{Source}
}
\startdata
\multicolumn{4}{l}{\bf Identifying Information} \\
\hline
TIC ID & 454140642 & & 1 \\
Gaia ID & 3255659981455492608 & & 2 \\
Tycho Reference ID & 0074-01254-1 & & 3 \\
$\alpha$  (J2000, hh:mm:ss) & 4:19:05.63 &  & 2 \\
$\delta$  (J2000, dd:mm:ss) & +00:54:00.2 &  & 2 \\
$\mu_{\alpha}$ (mas~yr$^{-1}$) & $2.539$ & 0.023 & 2 \\
$\mu_{\delta}$ (mas~yr$^{-1}$) & $-10.670$ & 0.018 & 2 \\
$\varpi$ (mas) & 2.7864 & 0.0215 & 2 \\
Distance (pc) & 358.8893 & 0.0407 & 2\\
\\
\multicolumn{4}{l}{\bf Photometric Properties} \\
\hline
$T$ (mag) & 9.8549 & 0.0097 & 1 \\
$B$ (mag) & 10.945 & 0.111 & 4\\
$V$ (mag) & 10.409 & 0.008 & 1 \\
$Gaia$ (mag) & 10.2224 &  0.0028 & 2 \\
$G_{BP}$ (mag) & 10.49970 & 0.001265 & 2 \\
$G_{RP}$ (mag) & 9.80662 & 0.001475 & 2 \\
$B_T$ (mag) & 10.769 & 0.058 & 3\\
$V_T$ (mag) & 10.329 & 0.063 & 3\\
$g'$ (mag) & 10.613 & 0.098 & 4\\
$r'$ (mag) & 10.229 & 0.041 & 4\\
$i'$ (mag) & 10.123 & 0.021 & 4\\
$J$ (mag) & 9.349 & 0.027 & 5 \\
$H$ (mag) & 9.116 & 0.026 & 5 \\
$K$ (mag) & 9.022 & 0.019 & 5 \\
$W1$ (mag) & 9.010 & 0.023 & 6 \\
$W2$ (mag) & 9.036 & 0.020 & 6 \\
$W3$ (mag) & 9.046 & 0.032 & 6 \\
$W4$ (mag) & 8.431 & 0.028 & 6 \\
$FUV$ (mag)& 18.381 & 0.195 & 7 \\
$NUV$ (mag)& 12.771 & 0.007 & 7 \\
Contamination & 0.0014 & & 1 \\
\\
\multicolumn{4}{l}{\bf Stellar Properties} \\
\hline
Teff, Aa (K) & 6434 & 30 & This work\\
Teff, Ab (K) & 6303 & 30 & This work\\
Teff, Ba (K) & 6303 & 30 & This work\\
Teff, Bb (K) & 6188 & 35 & This work\\
$\lbrack $Fe/H$ \rbrack$  & -0.039 & 0.022 & This work \\
Age (Gyr) & 1.945 & .27 & This work\\
\enddata
Sources: (1) TIC-8 \citep{TIC}, (2) Gaia EDR3 \citep{EDR3}, (3) Tycho-2 catalog \citep{Tycho2}, (4) APASS DR9 \citep{APASS9}, (5) 2MASS All-Sky Catalog of Point Sources \citep{2MASS}, (6) AllWISE catalog \citep{WISE}, (7) GALEX-DR5 (GR5) \citep{GALEX}

\tablenotetext{}{}
\end{deluxetable}

\section{Photometric data}
\label{sec:archival}
\begin{figure*}
    \centering
    \includegraphics[width=0.99\linewidth]{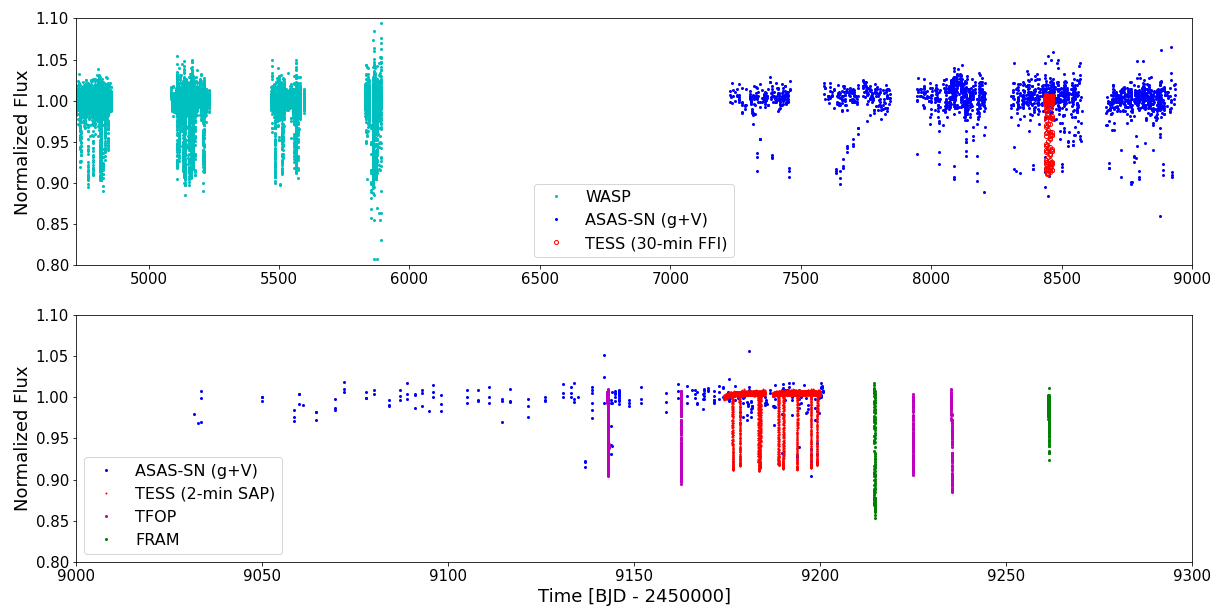}
    \caption{Photometry of the target from ASAS-SN (blue symbols for both g-filter and V-filter), FRAM (green symbols), TESS (red symbols), TFOP (magenta symbols), and WASP 200-mm lenses (cyan symbols), highlighting the covered baseline.}
    \label{fig:archival_ALL}
\end{figure*}

There is extensive archival data on \ticstar\ from the WASP and ASAS-SN surveys (\citealt{WASP2006}, \citealt{2014ApJ...788...48S}, \citealt{2017PASP..129j4502K}), altogether covering more than 4000 days. The time span of the available photometric data is shown in Figure~\ref{fig:archival_ALL}, which includes the TESS data as well as our follow-up photometry.  

\begin{figure*}
    \centering
    \includegraphics[width=0.45\linewidth]{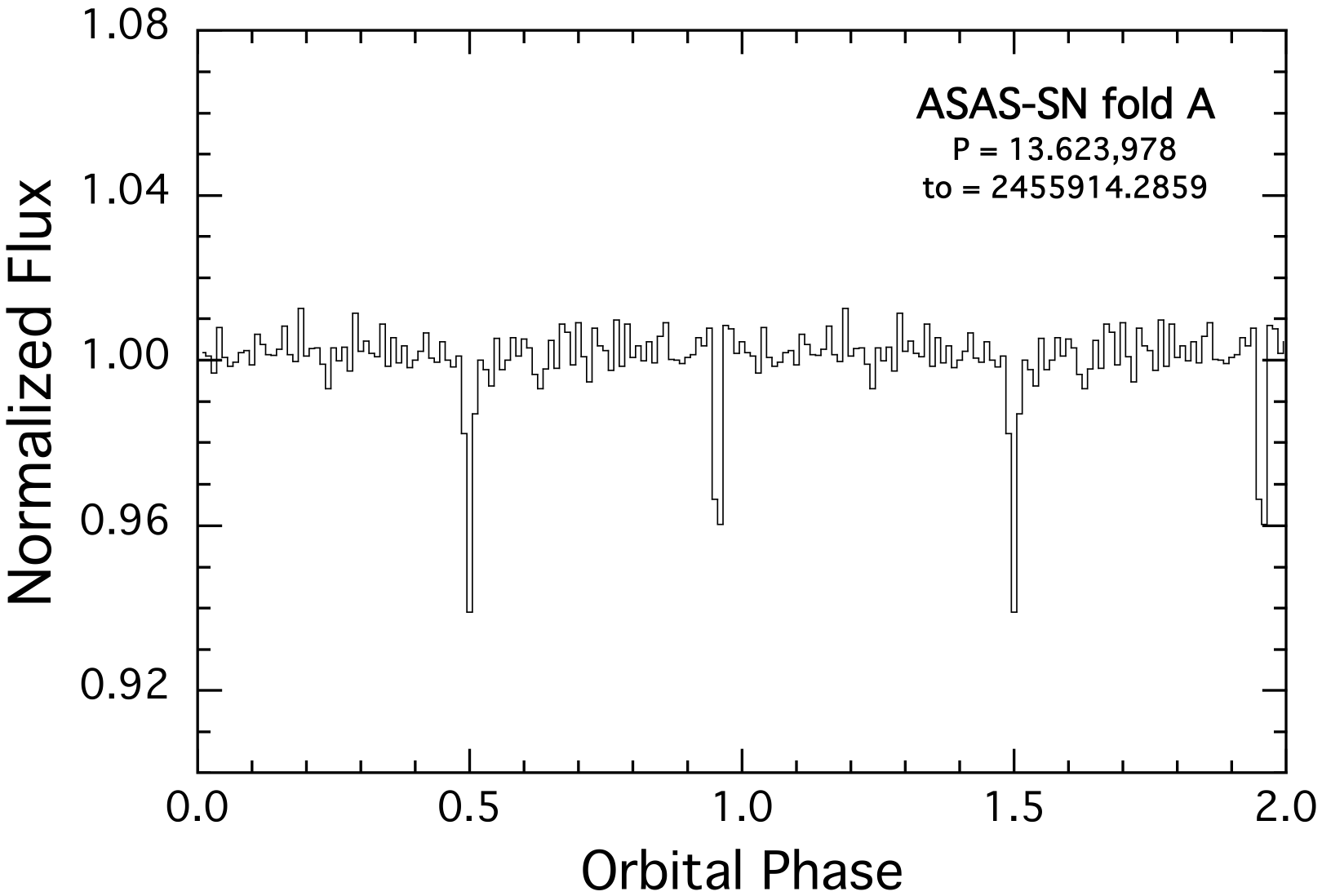}
    \includegraphics[width=0.45\linewidth]{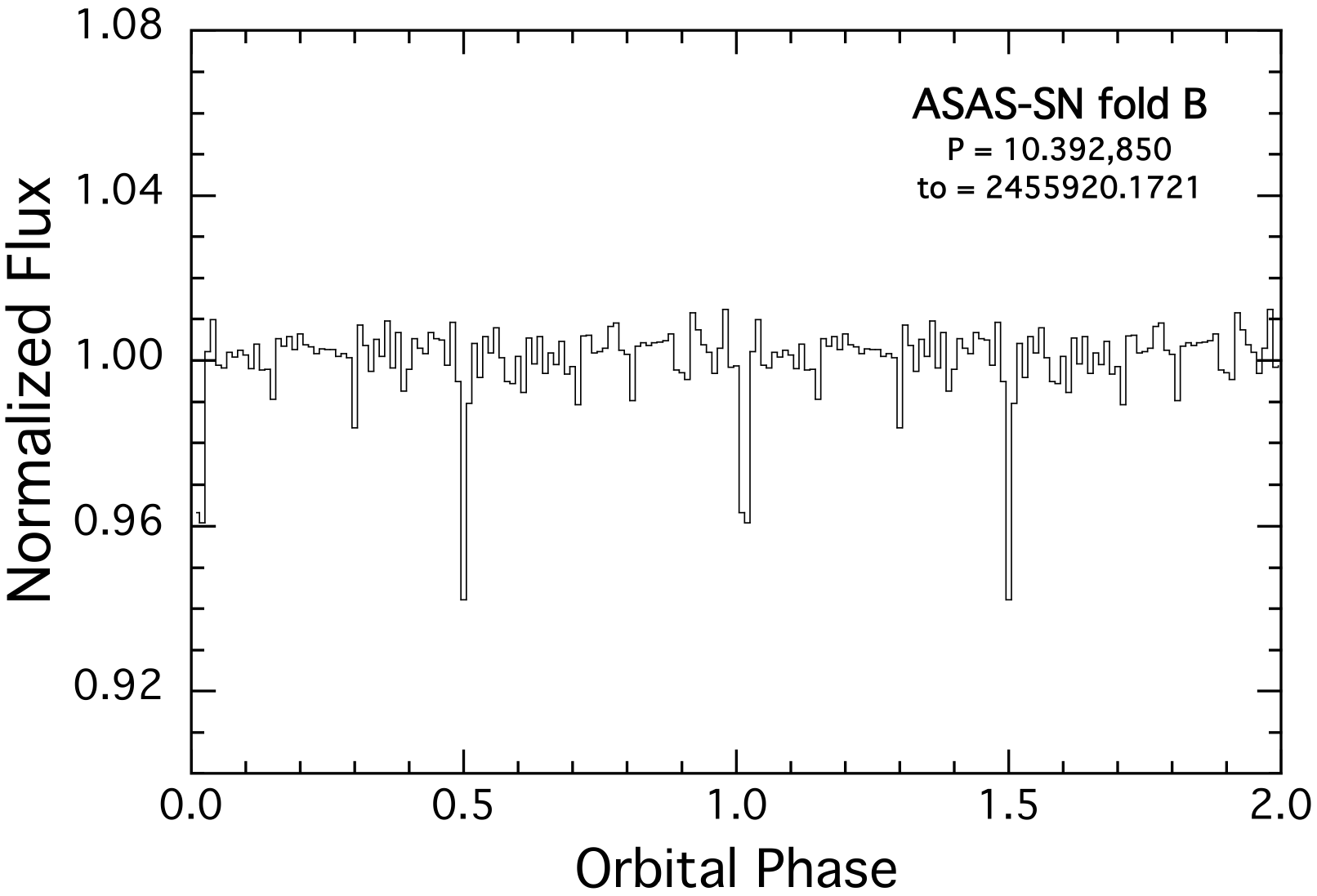}
    \includegraphics[width=0.45\linewidth]{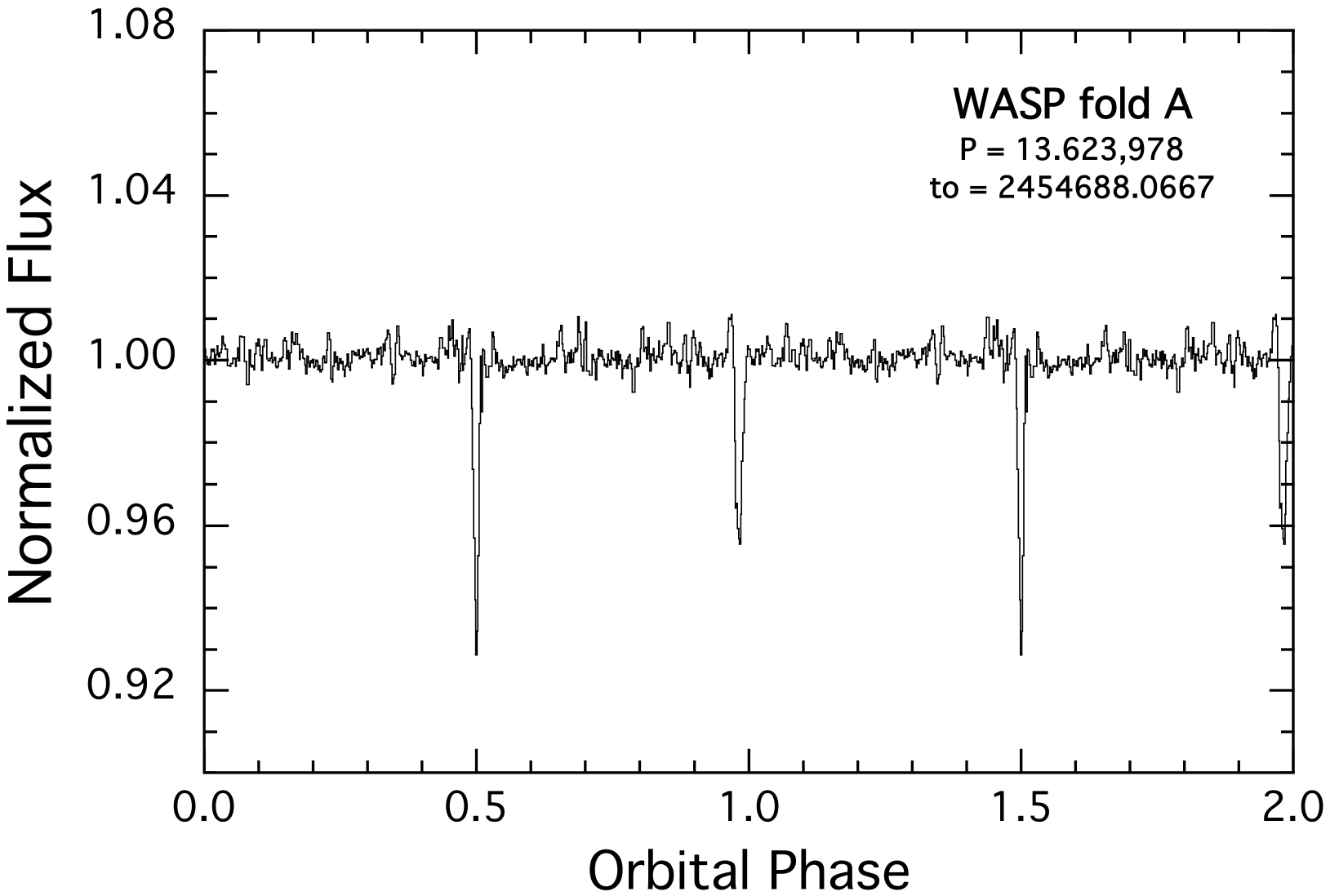}
    \includegraphics[width=0.45\linewidth]{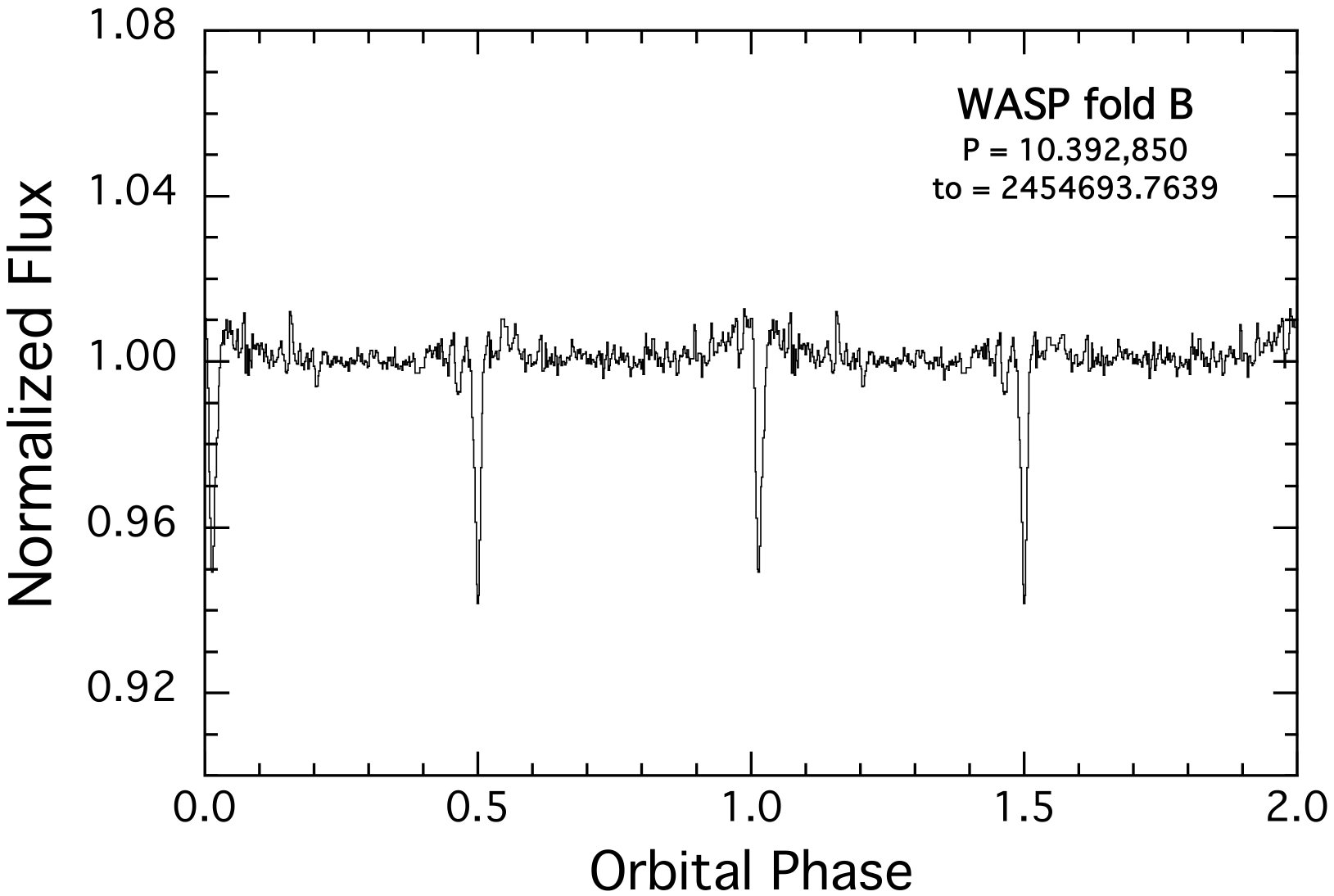}
    \caption{Folds of the ASAS-SN (top) and WASP (bottom) archival data for the two binaries about their respective orbital periods.}  
\label{fig:a_w_folds}
\end{figure*}

\begin{figure*}
    \centering
    \includegraphics[width=0.9\linewidth]{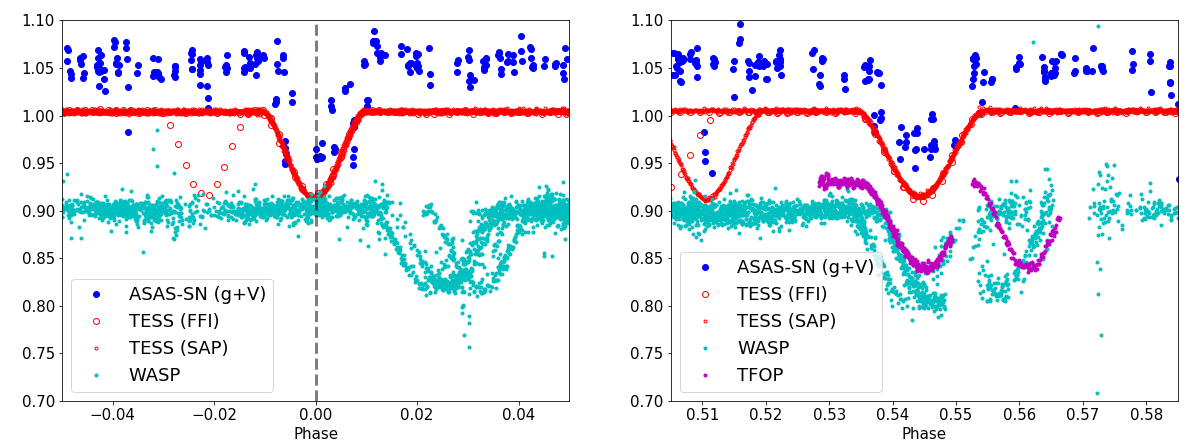}
    \includegraphics[width=0.9\linewidth]{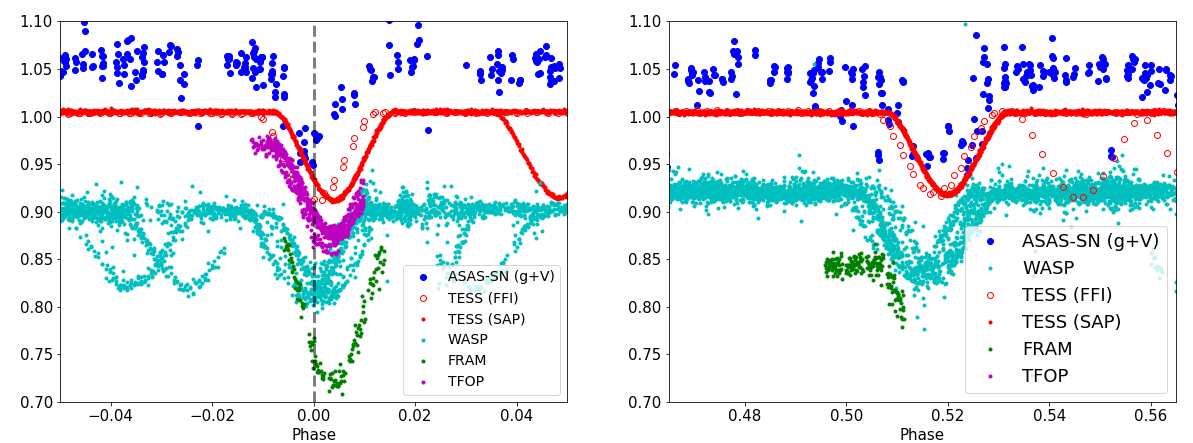}
    \caption{Upper panels: Phase-folded lightcurve of \ticstar\ for the primary (left) and secondary (right) eclipses of Binary A (using the global fitted ephemerides from Table \ref{tab:Fourier_ephemeris}), highlighting the prominent apsidal motion. The different datasets are vertically offset for clarity. Lower panels: Same as above but for Binary B. }
    \label{fig:folded_archival}
\end{figure*}

The ASAS-SN data of TIC 454140642 provide coverage of about 2000 days and overlap with TESS data. The target was observed by WASP with both the 85-mm and the 200-mm lenses, although the SNR of the former is poor and thus for our analysis we use only the latter. All four sets of eclipses are clearly recovered in the ASAS-SN and WASP data as shown in Figure \ref{fig:a_w_folds}. Box Least Squares" analysis (BLS, \citealt{BLS2002}) analysis of the two datasets provides similar orbital periods for the two binaries, as indicated in the figure captions and listed in Table \ref{tab:Fourier_ephemeris}. We note that the periods listed in the table are derived from the photometry alone; the final orbital periods are derived from the comprehensive spectro-photodynamical model described in Section \ref{sect:photdyn_anal}. Altogether, the available photometry phase-folded on the periods listed in Table \ref{tab:Fourier_ephemeris} shows clear apsidal motion (see Fig. \ref{fig:folded_archival}) and indicates dynamical interactions between the two binaries. 

\begin{table*}
\centering \caption{Derived ephemerides for the two EBs in TIC 454140642 from ASAS-SN, TESS, and WASP data.}
 \label{tab:Fourier_ephemeris}
 \begin{tabular}{c|cc|cc}
 \hline \hline
Binary               & \multicolumn{2}{c|}{A}       & \multicolumn{2}{c}{B} \\
\hline
${\bf ASAS-SN}$ & & & & \\
Period [days] & \multicolumn{2}{c|}{13.623978} &  \multicolumn{2}{c}{10.392850}\\
T0 [BJD - 2450000] & \multicolumn{2}{c|}{5914.2859} & \multicolumn{2}{c}{5920.1721}\\
\hline
${\bf TESS}$ & & & & \\
Period [days] & \multicolumn{2}{c|}{13.62395} &  \multicolumn{2}{c}{10.39335}\\
T0 [BJD - 2450000] &  \multicolumn{2}{c|}{8440.8394} &  \multicolumn{2}{c}{8445.5638}\\
\hline
${\bf WASP}$ & & & & \\
Period [days] & \multicolumn{2}{c|}{13.623978} & \multicolumn{2}{c}{10.392850}\\
T0 [BJD - 2450000] & \multicolumn{2}{c|}{4694.1464} & \multicolumn{2}{c}{4693.7639}\\
\hline
Global Fitted Periods & \multicolumn{2}{c|}{13.6239} & \multicolumn{2}{c}{10.3928}\\
Global Fitted T0 [BJD - 2450000] & 
\multicolumn{2}{c|}{8454.4688} & \multicolumn{2}{c}{8445.5610}\\
\hline
\hline
\end{tabular}
\end{table*}

\section{Follow-up observations}
\label{sec:follow-up}

\subsection{Spectroscopy}
\label{sec:spec}

\ticstar\ was monitored spectroscopically with the Tillinghast Reflector Echelle Spectrograph \citep[TRES;][]{Szentgyorgyi:2007, Furesz2008} mounted on the 1.5m Tillinghast reflector at the Fred L.\ Whipple Observatory on Mount Hopkins (AZ). TRES is a bench-mounted, fiber-fed instrument with a resolving power of $R \simeq 44,000$, and a wavelength coverage of 3900--9100~\AA\ in 51 orders. In total we gathered 26 spectra between September of 2020 and February of 2021, with signal-to-noise ratios in the region of the \ion{Mg}{1}\,b triplet ($\sim$5185~\AA) ranging from 28 to 60 per resolution element of 6.8~km s$^{-1}$. Exposure times ranged between 450 and 2400 seconds. The spectra were extracted and reduced as per \cite{Buchhave2010}, with wavelength solutions derived from bracketing thorium-argon lamp exposures.

\begin{figure}[h]
\centering
\includegraphics[width=0.45\textwidth]{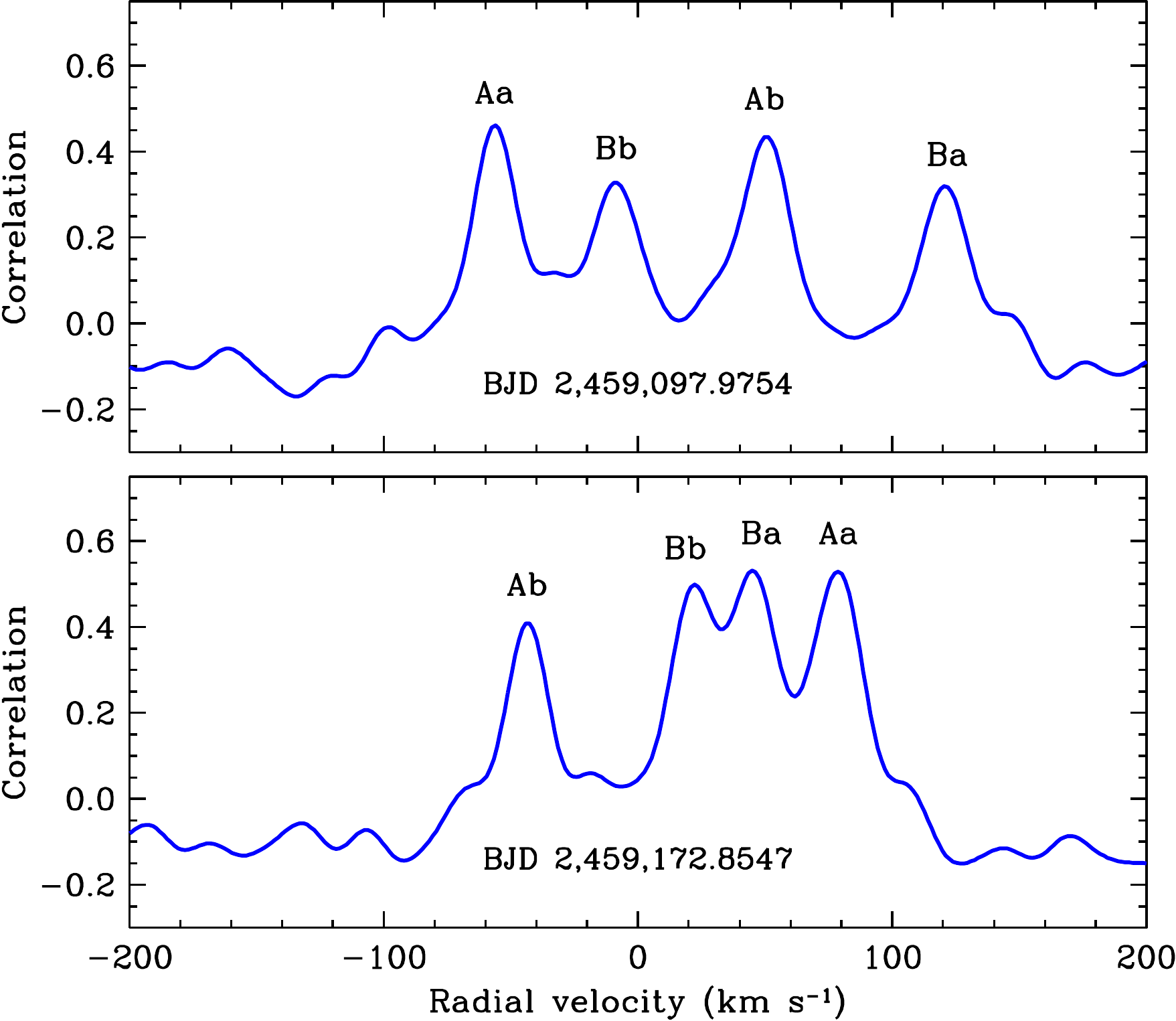}
\caption{Sample 1-D cross-correlation functions for \ticstar\ showing resolved peaks for the four components.}
\label{fig:ccf}
\end{figure}

The very first spectrum of \ticstar\ showed four sets of lines, which we expected should represent the two components of two double-lined binaries in the system (see Figure~\ref{fig:ccf}). Radial velocities were measured using {\tt QUADCOR}, a four-dimensional cross-correlation algorithm introduced by \cite{Torres:2007}, which uses four separate and possibly different templates matched to each star. In this case, however, the stars turn out to be quite similar to each other (see below), so we adopted the same template for the four components. We used a synthetic spectrum taken from a large library of calculated spectra based on model atmospheres by R.\ L.\ Kurucz \citep[see][]{Nordstrom:1994, Latham:2002}, covering the region of the \ion{Mg}{1}~b triplet. The template parameters were $T_{\rm eff} = 6250$~K, $\log g = 4.5$, ${\rm [m/H]} = 0.0$, and no rotational broadening. The choice of zero rotational broadening was based on experiments with a range of values for $v \sin i$, which indicated the broadening for all four stars is smaller than our spectral resolution. In addition to the velocities, we used {\tt QUADCOR} to measure the flux ratios among the components at the mean wavelength of our observations.

\begin{figure*}[h]
\centering
\includegraphics[width=0.85\textwidth]{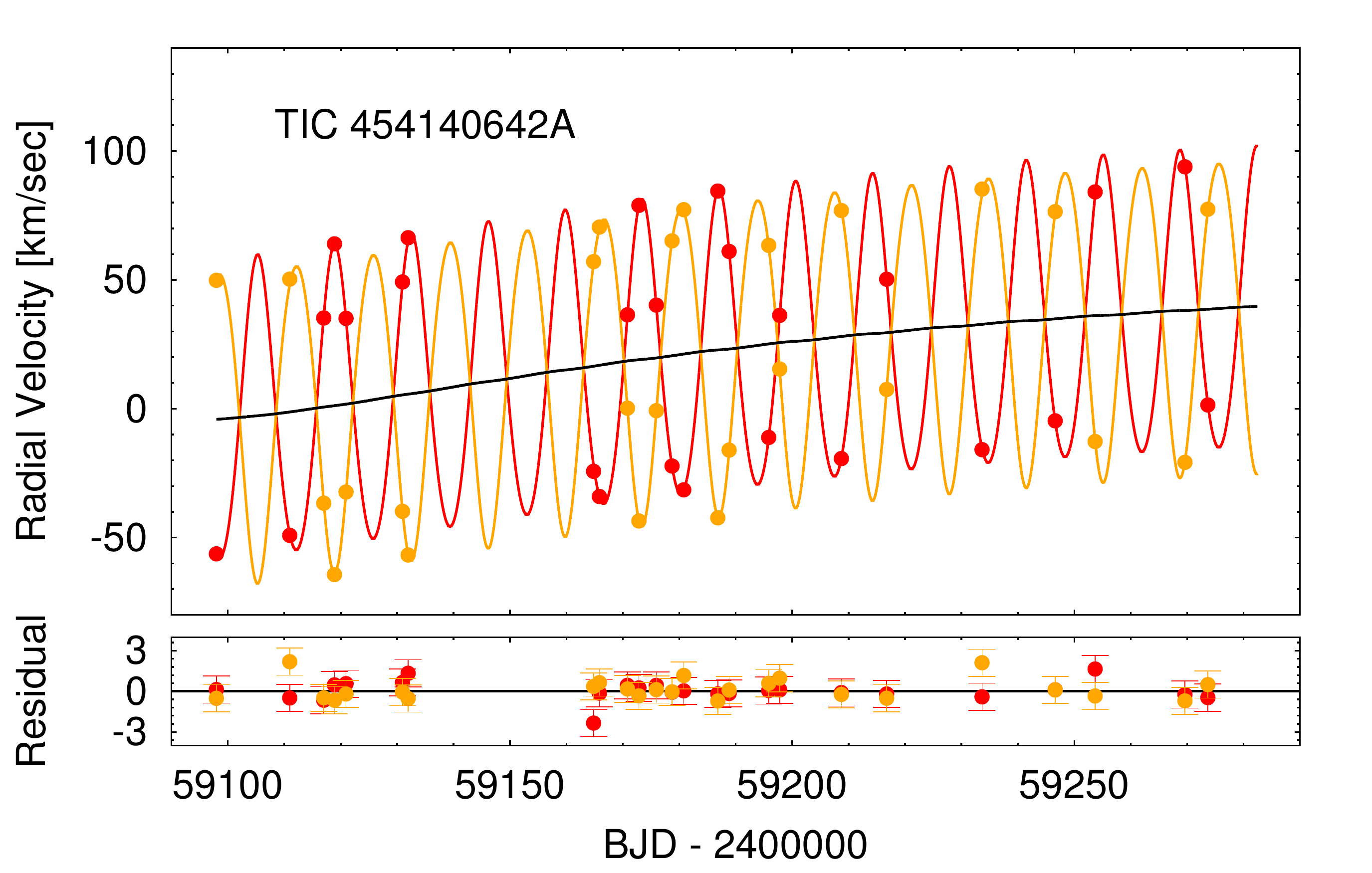}
\includegraphics[width=0.85\textwidth]{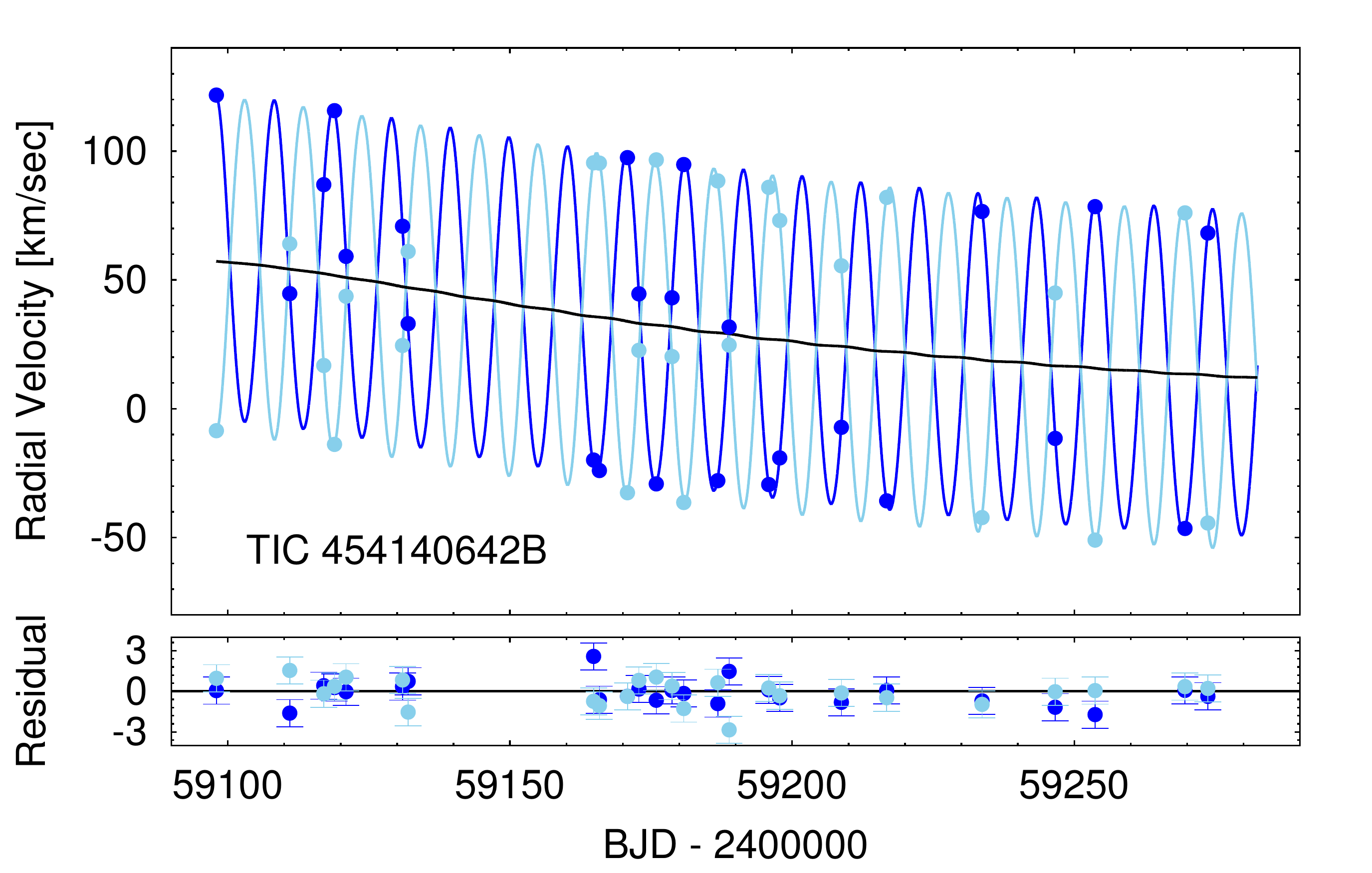}
\caption{Radial velocities measured for \ticstar\ (all individual RV values are given in Table \ref{tbl:RVs}). The solid curves represent the lowest $\chi^2$ spectro-photodynamical model solution (see in Sect. \ref{sect:photdyn_anal}). The black curves show the RV contributions of the outer orbit. The residuals are also presented in the bottom panels. Note that the data obtained after JD 2\,459\,200 were not used for the analysis, although they agree perfectly with the solution.}
\label{fig:RVs}
\end{figure*}

While approximate orbital periods were known from the photometry for the two binaries undergoing eclipses, at first it was not trivial to identify which set of lines corresponded to which component of which binary. In several of the spectra the lines were heavily blended, so the velocity determinations were not always possible for all four stars at this early stage, and were guided by the location of the peaks in plots of the one-dimensional cross-correlation functions. The process was further complicated by the degeneracy between the flux ratios and the velocity separations when the lines are blended, which made the velocities imprecise. Based on the more reliable flux ratios from the relatively few spectra with well separated lines, it was eventually realized that one star was brighter than the other three, another was the faintest, and the remaining two had similar brightness. Gathering the velocities for the brighter and fainter components, and sorting out the others based on the preliminary photometric ephemerides, we obtained initial orbital solutions for the two binaries based on the 18 observations through mid-December 2020. These fits showed considerable scatter, far in excess of the precision expected for the measurements. The primary and secondary residuals for the 13-day binary showed an obvious upward drift, and those of the 10-day binary a drift in the opposite direction (see Fig.~\ref{fig:RVs}). This was the first sure sign that the two binaries are gravitationally bound to each other, and in hindsight, explains our initial difficulty in identifying the components of the two binaries.

Solving for this linear drift separately in each of the double-lined orbits improved the velocities significantly, and allowed us to determine approximate binary elements. In addition to the period ($P$), the elements for each binary are the velocity semiamplitudes of the components ($K$), the orbital eccentricity ($e$) and argument of periastron of the secondary ($\omega$), the common center-of-mass velocity of the primary and secondary ($\gamma$), a reference time of periastron passage ($\tau$), and a radial acceleration term ($\dot{\gamma}$) for the binary center of mass. The orbital periods were better determined from the {\it TESS\/} photometry as well as the WASP and ASAS-SN archival data (see Section~\ref{sec:archival}), and were held fixed. These elements enabled us to better predict the velocities for the four components at epochs of heavy line blending, which were then used as a starting point for a refined analysis of those 18 observations and subsequent ones with {\tt QUADCOR}. We report our final velocities for all epochs in Table \ref{tbl:RVs}. Typical uncertainties for stars Aa, Ab, Ba, and Bb are 0.66, 0.73, 0.95, and 1.01~km s$^{-1}$, respectively, and are based on the rms scatter of the residuals from the orbits described above. The spectroscopic flux ratios we obtained for stars Ab, Ba, and Bb relative to the primary of the 13-day binary (Aa) are 0.85, 0.81, and 0.67 respectively, at the mean wavelength of our observations ($\sim$5185~\AA).

The first-cut binary parameters determined from this analysis are given in Table \ref{tbl:RVfit}, and served to set initial values for the global photodynamical analysis described in Section~\ref{sec:sysparms}.
The masses of the four stars in the quadruple system average 1.15~$M_{\sun}$ (assuming $\sin i \approx 1$, as both binaries are eclipsing), and none of the stars has a mass that differs from this average by more than 0.05~$M_{\sun}$. Given that the total masses of the two binaries are rather similar, we note also that the straight average of the $\gamma$ velocities, $(\gamma_{\rm A} + \gamma_{\rm B})/2 \simeq 26$~km s$^{-1}$, then reflects the approximate radial velocity of the system center of mass. Likewise, we expect that the radial accelerations for each binary will be nearly the same in magnitude, but of opposite sign --- which they indeed are: $\dot{\gamma}_A = +0.320 \pm 0.006$ cm s$^{-2}$, and $\dot{\gamma}_B = -0.331 \pm 0.008$ cm s$^{-2}$.

Finally, we used the relative velocities and accelerations to derive a preliminary estimate of the outer orbital period and separation under the assumptions that the outer quadruple orbit (i) is being viewed nearly edge-on (i.e., $i_{\rm out} \simeq 90^\circ$), and (ii) can be approximated as circular.  The former assumption is based on the fact that the EB orbits are both viewed nearly edge-on (but we caution that this is not a proof of the overall `flatness' of the system).  As for the assumption of a circular outer orbit, it is actually $e_{\rm out} \simeq 0.32$, but we proceeded under the circular orbit assumption to obtain a quick and simple estimate of the scope of the outer orbit.  There are two equations that can be solved to find the outer orbital separation, $a_{\rm out}$, and phase, $\phi_{\rm out}$, at the time of the RV observations, where $\phi$ is the angle from inferior conjunction of binary A:
\begin{eqnarray}
\sqrt{G(M_A+M_B)/a} \, \sin \phi & = & \gamma_A-\gamma_B \\
G(M_A+M_B)/a^2 \cos \phi & = & \dot{\gamma}_A - \dot{\gamma}_B
\end{eqnarray}
If we solve for $a_{\rm out}$ and $\phi_{\rm out}$, and then calculate the outer period, $P_{\rm out}$ we find:
$$ a_{\rm out} = 389 \pm 20 ~R_\odot $$
$$ \phi = -38.8^\circ \pm 5^\circ$$
$$P_{\rm out} \simeq 419 \pm 19~{\rm d}$$
where the uncertainties are due only to those associated with the measurements used in the calculations, and not to the actual eccentric nature of the outer orbit.

These approximations for the outer orbit served to guide the full photodynamical modeling described later.

\begin{table}
\centering
\caption{Radial Velocity Measurements for \ticstar$^a$}
\begin{tabular}{lllll}
\hline
\hline
  Observation Date  &   RV Aa & RV Ab & RV Ba & RV Bb  \\
 BJD $-$ 2,400,000       &    km s$^{-1}$  &  km s$^{-1}$  &    km s$^{-1}$ &  km s$^{-1}$    \\  
\hline
    59097.9754 &  $-$56.29 &   +49.83 &  +121.71 &   $-$8.53 \\
    59110.9829 &  $-$49.11 &   +50.30 &   +44.69 &   +64.00 \\
    59117.0083 &   +35.20 &  $-$36.63 &   +86.98 &   +16.78 \\
    59118.9106 &   +63.97 &  $-$64.35 &  +115.66 &  $-$13.81 \\
    59120.9626 &   +35.06 &  $-$32.31 &   +59.15 &   +43.65 \\
    59130.9687 &   +49.18 &  $-$39.79 &   +70.83 &   +24.54 \\
    59131.9596 &   +66.35 &  $-$56.74 &   +33.01 &   +61.01 \\
    59164.8201 &  $-$24.29 &   +57.10 &  $-$19.90 &   +95.43 \\
    59165.8458 &  $-$34.06 &   +70.49 &  $-$23.97 &   +95.35 \\
    59170.8216 &   +36.46 &    +0.21 &   +97.46 &  $-$32.57 \\
    59172.8547 &   +78.94 &  $-$43.53 &   +44.59 &   +22.68 \\
    59175.9399 &   +40.22 &   $-$0.75 &  $-$29.14 &   +96.57 \\
    59178.7227 &  $-$22.22 &   +65.17 &   +43.06 &   +20.25 \\
    59180.7983 &  $-$31.47 &   +77.27 &   +94.78 &  $-$36.39 \\
    59186.8346 &   +84.48 &  $-$42.34 &  $-$27.90 &   +88.42 \\
    59188.8468 &   +61.05 &  $-$16.05 &   +31.70 &   +24.74 \\
    59195.8602 &  $-$11.14 &   +63.39 &  $-$29.43 &   +85.94 \\
    59197.8081 &   +36.24 &   +15.42 &  $-$19.08 &   +73.06 \\
    59199.7751 &   +82.82  &  $-$32.63 &  +47.12  &  +4.03 \\
    59208.7251 &  $-$19.33 &   +76.92 &   $-$7.22 &   +55.46 \\
    59216.7606 &   +50.26 &    +7.48 &  $-$35.74 &   +82.01 \\
    59233.6625 &  $-$15.88 &   +85.25 &   +76.54 &  $-$42.23 \\
    59246.6163 &   $-$4.70 &   +76.49 &  $-$11.52 &   +44.92 \\
    59253.6860 &   +84.15 &  $-$12.72 &   +78.43 &  $-$50.96 \\
    59269.6177 &   +93.88 &  $-$20.79 &  $-$46.47 &   +76.01 \\
    59273.6654 &    +1.48 &   +77.41 &   +68.18 &  $-$44.34 \\
\hline 
\label{tbl:RVs} 
\end{tabular}

{Notes. (a) RV measurements were taken with TRES on the 1.5\,m reflector at the Fred Lawrence Whipple Observatory (FLWO) in Arizona.  See text for details.  The uncertainties on the individual measurements were determined empirically from the rms scatter in the residuals to the fitted model, and are 0.66, 0.73, 0.95, and 1.01~km s$^{-1}$\ for stars Aa, Ab, Ba, and Bb, respectively.} 

\end{table} 

\begin{table*}
\centering
\caption{Preliminary Orbital Solutions for \ticstar}
\begin{tabular}{lllll}
\hline
\hline
    Fitted Binary Parameters$^a$    & Aa & Ab & Ba & Bb \\ 
\hline
Period [days]$^b$ & 13.6239 & 13.6239 & 10.3928 & 10.3928 \\
$K$ [km s$^{-1}$] &  $57.89 \pm 0.41$  &   $60.79 \pm 0.40$  & $62.58 \pm 0.51$ &  $65.05 \pm 0.50$ \\ 
$e$ &  $0.075 \pm 0.005$  &    $0.075 \pm 0.005$ & $0.022 \pm 0.004$ &  $0.022 \pm 0.004$ \\ 
$\omega$ [deg]& $354.9 \pm 5$  &     $174.9 \pm 5$ & $1.9 \pm 16$ & $181.9 \pm 16$ \\ 
$\tau$$^c$ & $9159.9 \pm 0.2$   &   $9159.9 \pm 0.2$  & $9149.9 \pm 0.5$ & $9149.9 \pm 0.5$  \\ 
$\gamma$ [km s$^{-1}$] &   $+11.34 \pm 0.22$  & $+11.34 \pm 0.22$ & $+41.17 \pm 0.25$  & $+41.17 \pm 0.25$ \\ 
$\dot{\gamma}$ [cm s$^{-2}$] &    $0.320  \pm 0.006$ & $0.320  \pm 0.006$ & $-0.331 \pm 0.008$ & $-0.331 \pm 0.008$ \\ 
Mass [$M_\odot$]$^d$ & $1.208 \pm 0.018$ & $1.151 \pm 0.018$  & $1.141 \pm 0.019$ & $1.098 \pm 0.018$ \\
\hline 
\label{tbl:RVfit} 
\end{tabular}

{Notes. (a) Derived from the RV measurements alone.  (b) Fixed at the values given in Table \ref{tab:Fourier_ephemeris}. (c) BJD - 2450000. (d) Derived under the assumption that $\sin i \simeq 1$. } \end{table*}

\subsection{Photometric measurements}
\subsubsection{TESS Followup Observing Program}
We acquired ground-based time-series follow-up photometry of TIC 454140642 as part of the TESS Follow-up Observing Program (TFOP)\footnote{https://tess.mit.edu/followup}. We used the {\tt TESS Transit Finder}, which is a customized version of the {\tt Tapir} software package \citep{Jensen:2013}, to schedule our eclipse observations. The photometric data were extracted using {\tt AstroImageJ} \citep{Collins:2017}. Observations were acquired using the Las Cumbres Observatory network \citep[LCO;][]{Brown:2013} and the Hazelwood Private Observatory (Churchill, Victoria, Australia) as described in Table \ref{table:172900988sg1obs}.

\begin{table*}
\centering
\caption{Summary of Ground-based Photometric Observations from TFOP}
    \begin{tabular}{llllll}
    \hline\hline
    Telescope & Location & Date & Filter & Aperture radius & Coverage\\
           &           & [UTC]&        &   [arcsec]      & \\
    \hline

{\it  TIC 454140642 Primary Eclipse}\\
\hline
LCO-SAAO 0.4m  & South Africa   & 2020-07-18   &  $z$-short   & 5.7 & no event, ephemeris updated \\
LCO-McD 0.4m   & Texas, U.S.    & 2020-09-08   &  $z$-short   & 6.8 & no event, ephemeris updated  \\
LCO-Hal 0.4m   & Haleakala, U.S. & 2020-10-20   &  $z$-short   & 4.0 & ingress \\
LCO-SSO 0.4m  & Siding Spring   & 2021-01-10   &  $z$-short   & 5.1 & egress \\
Hazelwood 0.32m  & Australia   & 2021-01-10   &  $i^\prime$   & 5.0 & mid-eclipse \\
LCO-SAAO 0.4m  & South Africa   & 2021-01-20   &  $z$-short   & 4.0 & ingress \\[2mm] 

\hline

{\it  TIC 454140642 Secondary Eclipse}\\
\hline
LCO-Hal 0.4m   & Haleakala, U.S. & 2020-09-18   &  $z$-short   & 8.6 & no event, ephemeris updated \\
LCO-CTIO 0.4m  & South Africa   & 2020-11-09   &  $z$-short   & 6.8 & ingress \\[2mm] 

\hline
\end{tabular}
\label{table:172900988sg1obs}

{\em Notes.} The $z$-short filter is the same as the Pan-STARRS Z-short pass-band, i.e. ${\rm \lambda_{center} = 870~nm,~\lambda_{width} = 104~nm}$. The $i^\prime$ filter is the same as the SDSS $i^\prime$ pass-band, i.e. ${\rm \lambda_{center} = 755~nm,~\lambda_{width} = 129~nm}$.

\end{table*}

\subsubsection{Other follow-up ground based photometry}

In addition to the observations described above, we also used other photometric data for confirmation of the ETVs predicted from TESS data. These observations of eclipses of both pairs were obtained using the following instruments:
\begin{itemize}
    \item FRAM telescopes: 30-cm telescope as part of the Pierre Auger Observatory in Argentina \citep{2021arXiv210111602T},
    and 25-cm telescope in La Palma as part of the Cherenkov Telescope Array (CTA) \citep{2019arXiv190908085E}.
    \item Ond\v{r}ejov Observatory, Czech Republic, 65-cm reflecting telescope, using a G2-3200 CCD camera.
    \item Danish 1.54-m telescope located on La Silla in Chile, and using standard $V$ photometric filter.
    \item Small 34-mm refractor by R.U. - private observatory in J\'{\i}lov\'e u Prahy, Czech Republic, using a G2-0402 CCD camera.
\end{itemize}

\subsubsection{Speckle Imaging}

The target star was placed on the Southern Astrophysical Research Telescope (SOAR) speckle program in 2020 December. It was observed on 2021 February 27. The star was unresolved and no companions with $\rm \Delta I < 1\ mag$ and having a separation greater than $\rm 0.048\arcsec$ were detected. The achieved detection limits are $\rm \Delta I < 2.9\ mag$ at $\rm 0.15\arcsec$ and $\rm \Delta I < 4.4\ mag$ at $\rm 1\arcsec$. The details of these observations will be published jointly with other SOAR speckle results as part of the series of papers represented by \citet{Tokovinin2020}.  

The A-B pair, with an estimated semimajor axis of 5 mas, is not expected to be resolved at SOAR. However, the speckle imaging demonstrated the absence of other nearby faint stars that could otherwise be missed by the photometry or spectroscopy. To the best of our knowledge, then, \ticstar\ is an isolated quadruple system. 

\section{Analysis of System Parameters} 
\label{sec:sysparms}

\begin{table*}
\caption{Times of minima of TIC\,454140642 A}
 \label{Tab:TIC454140642A_ToM}
\begin{tabular}{@{}lrllrllrl}
\hline
BJD & Cycle  & std. dev. & BJD & Cycle  & std. dev. & BJD & Cycle  & std. dev. \\ 
$-2\,400\,000$ & no. &   \multicolumn{1}{c}{$(d)$} & $-2\,400\,000$ & no. &   \multicolumn{1}{c}{$(d)$} & $-2\,400\,000$ & no. &   \multicolumn{1}{c}{$(d)$} \\ 
\hline
54817.33051 & -267.0 & 0.00100 & 56227.60857 & -163.5 & 0.00016 & 59143.08895$^a$ & 50.5 & 0.00027 \\ 
54824.35102 & -266.5 & 0.01293 & 56268.46600 & -160.5 & 0.00019 & 59176.53298 & 53.0 & 0.00007 \\
55130.58298 & -244.0 & 0.00048 & 56574.59403 & -138.0 & 0.00025 & 59183.95313 & 53.5 & 0.00007 \\
55137.62361 & -243.5 & 0.00069 & 58440.84110 &   -1.0 & 0.00028 & 59190.15168 & 54.0 & 0.00006 \\
55171.45487 & -241.0 & 0.00033 & 58448.26119 &   -0.5 & 0.00039 & 59197.57368 & 54.5 & 0.00006 \\
55212.35313 & -238.0 & 0.00043 & 58454.45985 &    0.0 & 0.00040 & 59244.61950$^a$ & 58.0 & 0.00013 \\
55859.75174 & -190.5 & 0.00076 & 58461.88171 &    0.5 & 0.00045 &  &  &  \\
\hline		       
\end{tabular}

{\em Notes.} Integer and half-integer cycle numbers refer to primary and secondary eclipses, respectively. Eclipses between cycle nos. $-267.0$ and $-138.0$ were observed in data from the WASP project. Times of minima from cycle no. $-1.0$ to $0.5$ and $53.0$ to $54.5$ are determined from the TESS measurements. Times of minima labeled with $^a$ are obtained from ground-based follow-up observations.
\end{table*}

\begin{table*}
\caption{Times of minima of TIC\,454140642 B}
 \label{Tab:TIC454140642B_ToM}
\begin{tabular}{@{}lrllrllrl}
\hline
BJD & Cycle  & std. dev. & BJD & Cycle  & std. dev. & BJD & Cycle  & std. dev. \\ 
$-2\,400\,000$ & no. &   \multicolumn{1}{c}{$(d)$} & $-2\,400\,000$ & no. &   \multicolumn{1}{c}{$(d)$} & $-2\,400\,000$ & no. &   \multicolumn{1}{c}{$(d)$} \\ 
\hline
54766.52662 & -354.0 & 0.00173 & 55873.51828 & -247.5 & 0.00075 & 59162.70763$^a$ & 69.0 & 0.00016 \\ 
54813.46435 & -349.5 & 0.00057 & 56268.45172 & -209.5 & 0.00050 & 59178.45726 & 70.5 & 0.00038 \\
54818.50060 & -349.0 & 0.00025 & 56595.68135 & -178.0 & 0.00023 & 59183.49229 & 71.0 & 0.00004 \\
55109.43785 & -321.0 & 0.00372 & 56647.61573 & -173.0 & 0.00111 & 59188.84938 & 71.5 & 0.00033 \\
55135.59379 & -318.5 & 0.00024 & 56663.38911 & -171.5 & 0.00024 & 59193.88343 & 72.0 & 0.00004 \\
55156.38361 & -316.5 & 0.00040 & 58440.54817 &   -0.5 & 0.00004 & 59199.24128 & 72.5 & 0.00026 \\
55161.42771 & -316.0 & 0.00077 & 58445.56571 &    0.0 & 0.00007 & 59214.66803$^a$ & 74.0 & 0.00024 \\
55208.38458 & -311.5 & 0.00161 & 58455.95559 &    1.0 & 0.00006 & 59230.41337$^a$ & 75.5 & 0.00033 \\
55483.58635 & -285.0 & 0.00042 & 58461.32839 &    1.5 & 0.00012 & 59235.45419$^a$ & 76.0 & 0.00051 \\
\hline
\end{tabular}

{\em Notes.} Integer and half-integer cycle numbers refer to primary and secondary eclipses, respectively. Eclipses between cycle nos. $-354.0$ and $-171.5$ were observed in the WASP project. Times of minima from cycle no. $-0.5$ to $1.5$ and $70.5$ to $72.5$ are determined from the TESS measurements.  Times of minima labeled with $^a$ are obtained from ground-based follow-up observations.
\end{table*}

\subsection{Combined spectro-photodynamical analysis}
\label{sect:photdyn_anal}

We used the {\sc Lightcurvefactory} software package \citep{Borkovits2019,Borkovits2020} to carry out a simultaneous, joint analysis of the lightcurves, ETV, RV curves and multi-passband SED data of the system. We used three separate lightcurves as follows:
\begin{itemize}
\item[i)]{\textit{TESS} Sectors 5 and 32 lightcurves. In the case of the Sector 32 data we used the 2-min cadence dataset for extracting the ETV data; however, for the joint analysis we binned these data into 1800 sec, resulting an effective exposure length of 1425-sec, \footnote{As per the instrument handbook, the effective exposure time after cosmic-ray mitigation.} to get the same sampling as was available for Sector 5 data. Furthermore, we excluded the flat out-of-eclipse lightcurve sections and kept only the narrow $\pm0\fp03$-phase-domain regions around each eclipse.}
\item[ii)] Similarly clipped regions of archival WASP data. After the removal of outlier data points, these data were averaged into 1800-sec bins for the analysis.
\item[iii)] The ground-based eclipse observations of our follow-up photometric campaign. These data were also binned at 1800 sec.
\end{itemize}
Additionally, we used the four ETV curves (primary and secondary ETV data for both binaries; see Tables~\ref{Tab:TIC454140642A_ToM} and \ref{Tab:TIC454140642B_ToM}), and the first 18 epochs of the four TRES RV curves (see Table~\ref{tbl:RVs}). Furthermore, the observed passband magnitudes tabulated in Table~\ref{tab:EBparameters} were also used for the SED analysis.\footnote{Similar to our previous works, for the SED analysis we used a minimum uncertainty of $0.03$ mag for most of the observed passband magnitudes, in order to avoid the outsized contribution of the extremely precise Gaia magnitudes and also to counterbalance the uncertainties inherent in our interpolation method during the calculations of theoretical passband magnitudes that are part of the fitting process. The two exceptions are the WISE $W4$ and GALEX $FUV$ magnitudes, for which the uncertainties were inflated to 0.3 mag.}

The modelling runs proceeded as follows. {\sc Lightcurvefactory} calculates the positions and velocities of each star at any requested instant via the numerical integration of the orbital motion. Then, taking into account the radii, temperatures and other atmospheric properties of each star, and their relative geometry with respect to the others and to the observer, the code predicts the net lightcurve of the whole system, enabling any kind of mutual (or multiple) eclipses. Note that in the present analysis, stellar radii, and temperatures were constrained through the stellar masses and the metallicity and age of the system with the use of built-in \texttt{PARSEC} stellar isochrones \citep{2012MNRAS.427..127B}, as was described in \citet{Borkovits2020} in detail. Furthermore, the same \texttt{PARSEC} tables are used for generating theoretical passband magnitudes for the SED fitting part of the analysis. To solve the inverse problem, {\sc Lightcurvefactory} employs a Markov Chain Monte Carlo (MCMC)-based parameter search, implementing the generic Metropolis-Hastings algorithm \citep[see e.g.][]{2005AJ....129.1706F}.

In most of the MCMC runs we adjusted the following 20 parameters:
\begin{itemize}
\item[(i)] 4 orbital elements for binary A, i.e., the eccentricity and argument of periastron in the combinations of $(e\sin\omega)_\mathrm{A}$ and $(e\cos\omega)_\mathrm{A}$, its inclination, $i_\mathrm{A}$ and the longitude of the ascending node $\Omega_\mathrm{A}$ relative to the node of binary B;
\item[(ii)] 3 orbital elements for binary B, i.e., the same as above with the exception of $\Omega_\mathrm{B}$, for which the value was kept at zero;
\item[(iii)] 6 orbital parameters for the outer orbit, as its anomalistic period, $P_\mathrm{AB}$, eccentricity and argument of periastron, inclination, periastron passage time, $\tau_\mathrm{AB}$ and, the longitude of the ascending node relative to the node of binary B;
\item[(iv)] 4 mass-related parameters, as the masses of the two primaries, $m_\mathrm{Aa,Ba}$), and the (inner) mass ratios of the two binaries, $q_\mathrm{A,B}$;
\item[(v)] 3 global parameters, as the (logarithmic) age of the system, its metallicity,  [m/H], and the interstellar reddening $E(B-V)$.
\end{itemize}
Moreover, the following parameters were internally constrained:
\begin{itemize}
\item[(i)] 4 remaining orbital parameters of the inner orbits, i.e., their periods ($P_\mathrm{A,B}$) and inferior conjunction times ($\mathcal{T}_\mathrm{A,B}^\mathrm{inf}$) at epoch $t_0$, which were constrained through the ETV curves \citep[see][]{Borkovits2019};
\item[(ii)] The systemic radial velocity of the entire quadruple system ($\gamma$) was obtained at each step by minimizing the $\chi^2_\mathrm{RV}$ value a posteriori;
\item[(iii)] 8 fundamental stellar parameters, i.e., the radii and temperatures of the four stars, which were interpolated from the \texttt{PARSEC} grids with the use of \{mass, metallicity, age\} triplets in the manner described by \citet{Borkovits2020};
\item[(iv)] The (photometric) distance of the system was also obtained through an a posteriori minimization of $\chi^2_\mathrm{SED}$;
\item[(v)] Finally, the (logarithmic) limb-darkening coefficients were calculated at each trial step internally from the pre-computed passband-dependent tables downloaded from the Phoebe 1.0 Legacy page\footnote{\url{http://phoebe-project.org/1.0/download}}. These tables  were originally used in former versions of the {\sc Phoebe} software \citep{2005ApJ...628..426P}.
\end{itemize}

\begin{figure}
\begin{center}
\includegraphics[width=0.49 \textwidth]{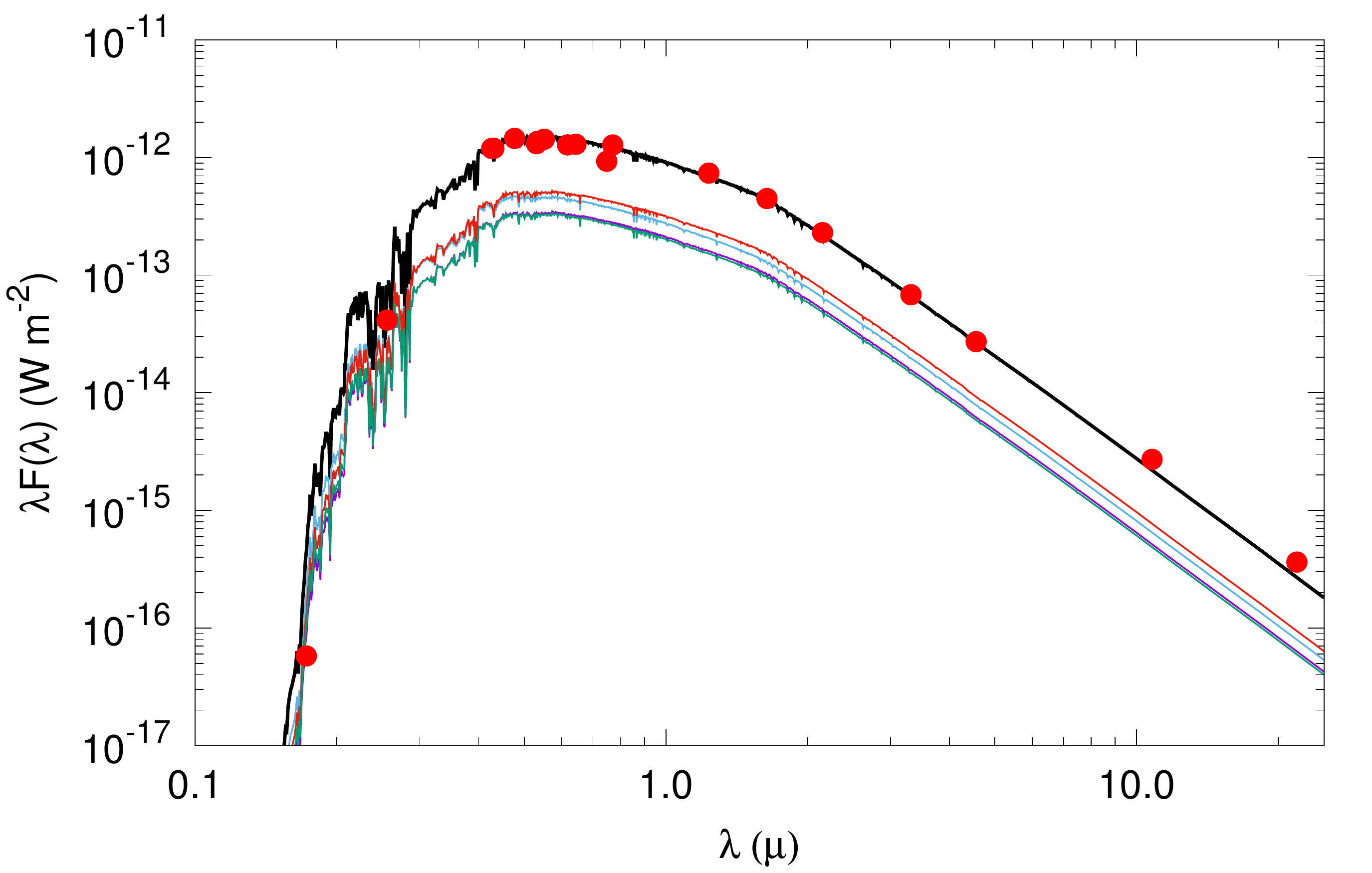} 
\caption{The summed SED of the four stars of TIC\,454140642. The dereddened observed magnitudes are converted into the flux domain (red filled circles), and overplotted with the quasi-continuous summed SED for the quadruple star system (thick black line). This SED is computed from the \citet{castelli03} ATLAS9 stellar atmospheres models (\url{http://wwwuser.oats.inaf.it/castelli/grids/gridp00k2odfnew/fp00k2tab.html}). The separate SEDs of the four stars are also shown with thin green, black and purple lines, respectively. } 
\label{fig:sedfit} 
\end{center}
\end{figure}

\begin{figure*}
\centering
\includegraphics[width=0.49\linewidth]{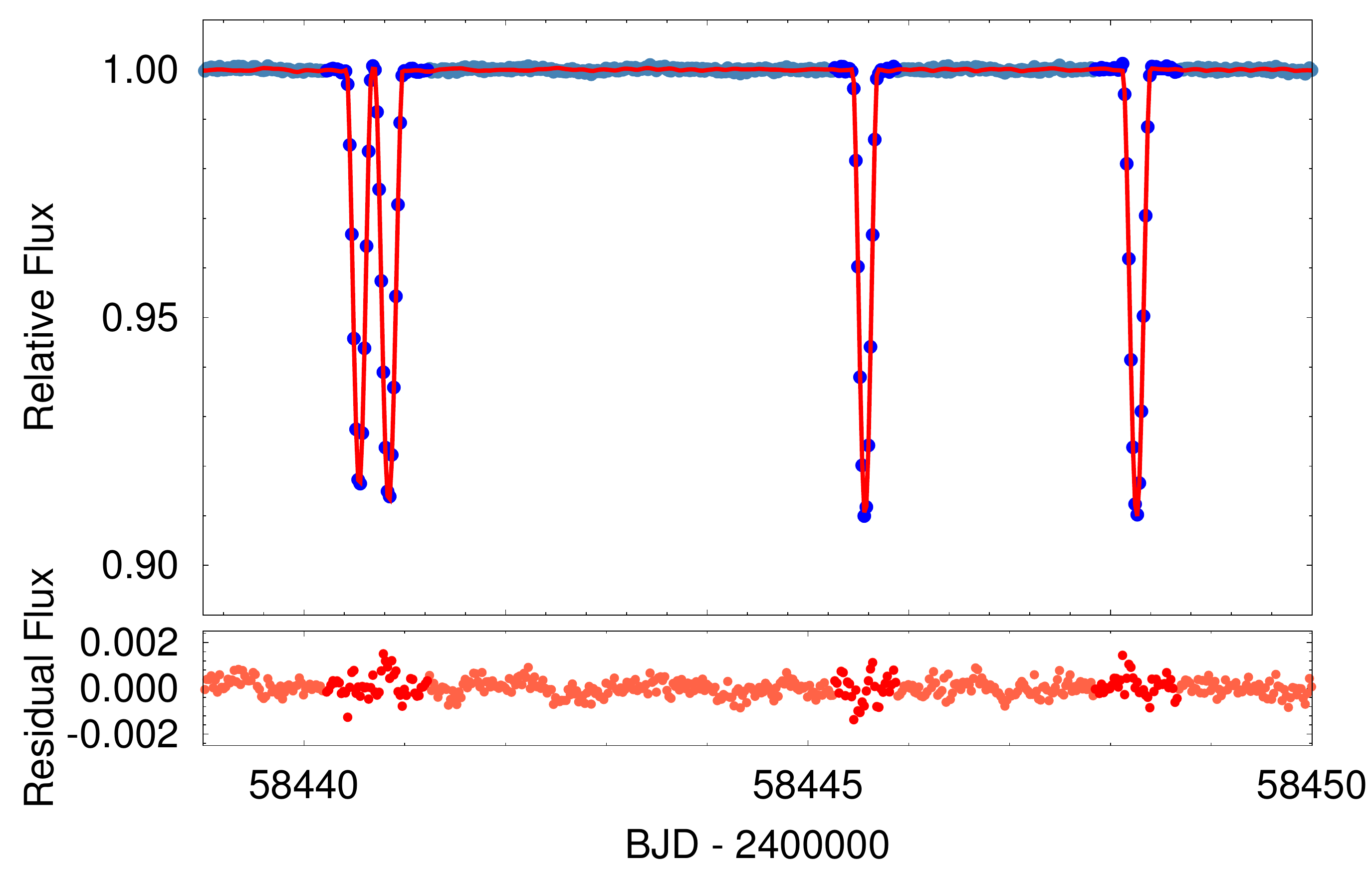}
\includegraphics[width=0.49\linewidth]{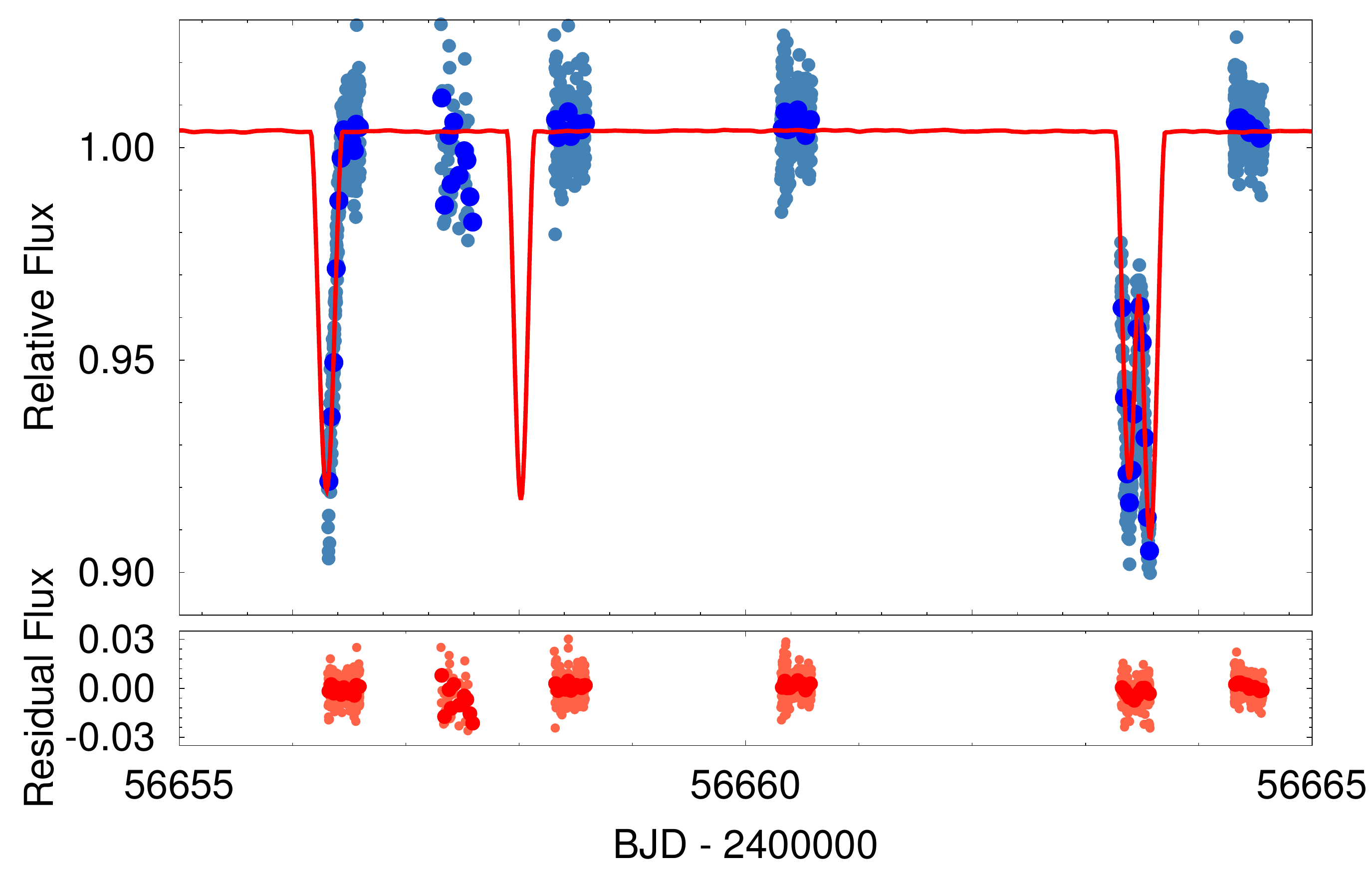}
\caption{Photodynamical model (red line) plotted with the TESS data (left, blue) and WASP data (right, blue). Dark blue dots represent the data points used for fitting the photodynamical model. In the case of \textit{TESS} data (left) these are the observed data points within the $\pm0\fp03$ phase-domain of each eclipse, while in the case of the WASP observations (right) these are the 1800-sec averages of the individual data points. The pale blue points represent the remaining, out-of-eclipse \textit{TESS} data (in the left panel), and the original WASP data (to the right). Residual curves are also plotted in the lower panels.}
 \label{fig:photo} 
\end{figure*}

Table~\ref{tbl:simlightcurve} lists the median values of the stellar and orbital parameters of the quadruple systems that are being either adjusted, internally constrained, or derived from the MCMC posteriors, together with their $1\sigma$ statistical uncertainties. The RV curves, lightcurves, ETV curves, and SED model of the lowest $\chi^2_\mathrm{global}$ solution are plotted in Figs.~\ref{fig:RVs}--\ref{fig:photo}.

\begin{table*}
\centering
\caption{Median values of the parameters from the double EB simultaneous lightcurve, $2\times$ SB2 radial velocity, double ETV, joint SED and \texttt{PARSEC} 
evolutionary track solution from {\sc Lightcurvefactory.}}
\begin{tabular}{lccccc}
\hline
\multicolumn{6}{c}{Orbital elements$^a$} \\
\hline
   & \multicolumn{3}{c}{subsystem}  \\
   & \multicolumn{2}{c}{A} & \multicolumn{2}{c}{B} & A--B \\
  \hline
$P_\mathrm{a}$ [days]            & \multicolumn{2}{c}{$13.61594_{-0.00016}^{+0.00016}$} & \multicolumn{2}{c}{$10.39063_{-0.00008}^{+0.00008}$}       & $432.1_{-0.4}^{+0.5}$ \\
semimajor axis  [$R_\odot$]      & \multicolumn{2}{c}{$31.87_{-0.08}^{+0.12}$}          & \multicolumn{2}{c}{$26.27_{-0.06}^{+0.07}$}                & $400.0_{-0.8}^{+1.0}$ \\  
$i$ [deg]                        & \multicolumn{2}{c}{$87.55_{-0.06}^{+0.06}$}          & \multicolumn{2}{c}{$87.52_{-0.08}^{+0.09}$}                & $87.69_{-0.47}^{+0.46}$  \\
$e$                              & \multicolumn{2}{c}{$0.07449_{-0.00021}^{+0.00021}$}  & \multicolumn{2}{c}{$0.02700_{-0.00015}^{+0.00022}$}        & $0.32285_{-0.00497}^{+0.00501}$ \\  
$\omega$ [deg]                   & \multicolumn{2}{c}{$161.21_{-0.56}^{+0.65}$}         & \multicolumn{2}{c}{$190.89_{-1.19}^{+1.74}$}               & $325.77_{-1.18}^{+1.28}$ \\
$\tau^b$ [BJD]     & \multicolumn{2}{c}{$2\,458\,437.0411_{-0.0204}^{+0.0240}$}           & \multicolumn{2}{c}{$2\,458\,443.3691_{-0.0342}^{+0.0502}$} & $2\,459\,070.05_{-1.25}^{+1.26}$\\
$\Omega$ [deg]                   & \multicolumn{2}{c}{$-0.13_{-0.28}^{+0.29}$}          & \multicolumn{2}{c}{$0.0$}                                  & $-0.15_{-0.26}^{+0.26}$ \\
$(i_\mathrm{m})_{A-...}^c$ [deg]   & \multicolumn{2}{c}{$0.0$}                            & \multicolumn{2}{c}{$0.25_{-0.13}^{+0.20}$}                 & $0.37_{-0.21}^{+0.34}$ \\
$(i_\mathrm{m})_{B-...}$ [deg]   & \multicolumn{2}{c}{$0.25_{-0.13}^{+0.20}$}           & \multicolumn{2}{c}{$0.0$}                                  & $0.47_{-0.24}^{+0.32}$ \\
$\varpi_\mathrm{dyn}^d$ [deg      ]& \multicolumn{2}{c}{$11.04_{-1.35}^{+1.35}$} & \multicolumn{2}{c}{$161.21_{-0.55}^{+0.55}$} & $325.82_{-1.19}^{+1.19}$ \\
$i_\mathrm{dyn}^d$ [deg] & \multicolumn{2}{c}{$0.31_{-0.17}^{+0.28}$} & \multicolumn{2}{c}{$0.41_{-0.21}^{+0.27}$} & $0.06_{-0.04}^{+0.06}$ \\
$i_\mathrm{inv}^e$ [deg] & \multicolumn{5}{c}{$87.67_{-0.40}^{+0.38}$} \\
$\Omega_\mathrm{inv}^e$ [deg] & \multicolumn{5}{c}{$-0.13_{-0.24}^{+0.24}$} \\
\hline
\multicolumn{6}{c}{RV curve related parameters} \\
\hline
mass ratio $[q=m_\mathrm{sec}/m_\mathrm{pri}]$ & \multicolumn{2}{c}{$0.959_{-0.006}^{+0.005}$} & \multicolumn{2}{c}{$0.964_{-0.005}^{+0.005}$} & $0.963_{-0.013}^{+0.009}$ \\
$K_\mathrm{pri}$ [km\,s$^{-1}$] & \multicolumn{2}{c}{$58.11_{-0.28}^{+0.24}$} & \multicolumn{2}{c}{$62.80_{-0.25}^{+0.24}$} & $24.25_{-0.13}^{+0.13}$ \\ 
$K_\mathrm{sec}$ [km\,s$^{-1}$] & \multicolumn{2}{c}{$60.62_{-0.20}^{+0.25}$} & \multicolumn{2}{c}{$65.14_{-0.21}^{+0.21}$} & $25.20_{-0.16}^{+0.21}$ \\ 
$\gamma$ [km/s]                 & \multicolumn{2}{c}{$-$} & \multicolumn{2}{c}{$-$} & $25.97_{-0.08}^{+0.07}$\\  
  \hline  
\multicolumn{6}{c}{Apsidal motion related parameters$^f$} \\  
\hline  
$U_\mathrm{num}$ [year] & \multicolumn{2}{c}{$70$} & \multicolumn{2}{c}{$97$} & $472$\\
$U_\mathrm{theo}$ [year] & \multicolumn{2}{c}{$87_{-1}^{+1}$} & \multicolumn{2}{c}{$109_{-1}^{+1}$} & $480_{-5}^{+5}$\\
$\Delta\omega_\mathrm{3b}$ [arcsec/cycle] & \multicolumn{2}{c}{$556_{-5}^{+4}$} & \multicolumn{2}{c}{$338_{-3}^{+3}$} & $3194_{-32}^{+31}$\\
$\Delta\omega_\mathrm{GR}$ [arcsec/cycle] & \multicolumn{2}{c}{$0.611_{-0.003}^{+0.005}$} & \multicolumn{2}{c}{$0.709_{-0.003}^{+0.004}$} & $0.106_{-0.001}^{+0.001}$\\
$\Delta\omega_\mathrm{tide}$ [arcsec/cycle] & \multicolumn{2}{c}{$0.075_{-0.005}^{+0.004}$} & \multicolumn{2}{c}{$0.130_{-0.007}^{+0.009}$} & $-$\\
\hline
\multicolumn{6}{c}{Stellar parameters} \\
\hline
   & Aa & Ab &  Ba & Bb & \\
  \hline
 \multicolumn{6}{c}{Relative quantities} \\
  \hline
fractional radius [$R/a$]               & $0.0394_{-0.0006}^{+0.0005}$ & $0.0365_{-0.0005}^{+0.0004}$  & $0.0442_{-0.0005}^{+0.0007}$ & $0.0415_{-0.0005}^{+0.0005}$ & \\
fractional flux [in \textit{TESS}-band] & $0.305_{-0.008}^{+0.009}$    & $0.246_{-0.008}^{+0.007}$     & $0.247_{-0.009}^{+0.006}$    & $0.204_{-0.008}^{+0.006}$    & \\
fractional flux [in \textit{SWASP}-band]& $0.311_{-0.008}^{+0.010}$    & $0.245_{-0.009}^{+0.008}$     & $0.246_{-0.010}^{+0.007}$    & $0.200_{-0.009}^{+0.007}$    & \\
fractional flux [in $R_C$-band]         & $0.307_{-0.008}^{+0.009}$    & $0.245_{-0.008}^{+0.007}$     & $0.246_{-0.009}^{+0.007}$    & $0.202_{-0.008}^{+0.007}$    & \\
 \hline
 \multicolumn{6}{c}{Physical Quantities} \\
  \hline 
 $m$ [M$_\odot$]   & $1.195_{-0.009}^{+0.012}$ & $1.146_{-0.012}^{+0.013}$ & $1.146_{-0.007}^{+0.010}$ & $1.105_{-0.008}^{+0.010}$ & \\
 $R^g$ [R$_\odot$] & $1.255_{-0.019}^{+0.018}$ & $1.161_{-0.016}^{+0.018}$ & $1.161_{-0.015}^{+0.019}$ & $1.091_{-0.016}^{+0.015}$ & \\
 $T_\mathrm{eff}^g$ [K]& $6434_{-29}^{+31}$    & $6303_{-29}^{+31}$        & $6303_{-29}^{+33}$        & $6188_{-36}^{+34}$        & \\
 $L_\mathrm{bol}^g$ [L$_\odot$] & $2.420_{-0.090}^{+0.098}$ & $1.917_{-0.081}^{+0.079}$ & $1.917_{-0.082}^{+0.085}$ & $1.570_{-0.077}^{+0.070}$ &\\
 $M_\mathrm{bol}^g$ & $3.81_{-0.04}^{+0.04}$    & $4.06_{-0.04}^{+0.05}$    & $4.06_{-0.05}^{+0.05}$    & $4.28_{-0.05}^{+0.05}$    &\\
 $M_V^g           $ & $3.81_{-0.04}^{+0.04}$    & $4.07_{-0.05}^{+0.05}$    & $4.07_{-0.05}^{+0.05}$    & $4.29_{-0.05}^{+0.06}$    &\\
 $\log g^g$ [dex]   & $4.316_{-0.010}^{+0.013}$ & $4.365_{-0.009}^{+0.011}$ & $4.367_{-0.014}^{+0.010}$ & $4.405_{-0.011}^{+0.010}$ &\\
 \hline
\multicolumn{6}{c}{Global Quantities} \\
\hline
$\log$(age)$^g$ [dex] &\multicolumn{5}{c}{$9.289_{-0.064}^{+0.045}$} \\
$ [M/H]^g$  [dex]      &\multicolumn{5}{c}{$-0.039_{-0.023}^{+0.022}$} \\
$E(B-V)$ [mag]    &\multicolumn{5}{c}{$0.012_{-0.009}^{+0.013}$} \\
$(M_V)_\mathrm{tot}^g$           &\multicolumn{5}{c}{$2.54_{-0.03}^{+0.03}$} \\
distance [pc]                &\multicolumn{5}{c}{$357_{-5}^{+4}$}  \\  
\hline
\end{tabular}
\label{tbl:simlightcurve}

{\em Notes.} (a) Instantaneous, osculating orbital elements at epoch $t_0=2\,458437.5$; (b) Time of periastron passsage; (c) Mutual (relative) inclination; (d) Longitude of pericenter ($\varpi_\mathrm{dyn}$) and inclination ($i_\mathrm{dyn}$) with respect to the dynamical (relative) reference frame (see text for details); (e) Inclination ($i_\mathrm{inv}$) and node ($\Omega_\mathrm{inv}$) of the invariable plane to the sky; (f) See Sect~\ref{sect:orbitaldynamics} for a detailed discussion of the tabulated apsidal motion parameters; (g) Interpolated from the \texttt{PARSEC} isochrones;  
\end{table*}

In addition to the usual orbital elements tabulated in the first section of Table~\ref{tbl:simlightcurve}, we list the mutual (i.e.~relative to each other) inclinations of the three orbital planes as well (${i_\mathrm{mut}}_{A-B; A-AB; B-AB}$). We also provide the angular orbital elements in the dynamical frame of reference, i.e. the inclinations of the orbits relative to the invariable plane of the quadruple system ($i_\mathrm{dyn}$), and the longitude of the pericenter ($\varpi_\mathrm{dyn}=\Omega_\mathrm{dyn}+\omega_\mathrm{dyn}$) of each orbit defined as the sum of the dynamical longitude of the node\footnote{The angular distance of the intersection of the given orbital plane and the invariable plane -- the dynamical nodal line -- from the intersection of the plane of the sky and the invariable plane, measured along the invariable plane.} and the dynamical argument of pericenter\footnote{The angular distance of the pericenter from the dynamical ascending node along the orbital plane}. We note that due to the almost perfect coplanarity of the entire system, the intersections of the three orbits with the invariable plane (and therefore all the three pairs of $\Omega_\mathrm{dyn}$ and $\omega_\mathrm{dyn}$) become ill-defined; however, their sums, i.e. $\varpi_\mathrm{dyn}=\Omega_\mathrm{dyn}+\omega_\mathrm{dyn}$ remain well-defined and we tabulate the latter quantities in Table~\ref{tbl:simlightcurve}. The relevance of these dynamical orbital elements lies in the fact that they, along with the mutual inclinations, appear directly in the analytic perturbation theories of multiple star systems \citep[see, e.g.][for further discussions]{Borkovits2015}. Furthermore, in this section we tabulate also some apsidal motion related parameters which will be discussed in Sect.~\ref{sect:orbitaldynamics}. Finally, for completeness, we also provide the inclination and nodal angle of the invariable plane ($i_\mathrm{inv}$, $\Omega_\mathrm{inv}$, respectively).

We note that the values listed in Table~\ref{tbl:simlightcurve} represent the (instantaneous) osculating orbital elements at the (arbitrarily chosen) epoch $t_0=2\,458437.5$. Therefore, the respective periods listed in the first row of the table should not be used for eclipse timing prediction for future follow-up observations. This is illustrated in Fig.~\ref{fig:periods} where we show the various relevant periods for the system. The osculating orbital elements (grey lines in the figure) are calculated at each instant from the Cartesian (Jacobian) coordinates and velocities of each orbit as if they represented pure, unperturbed, Keplerian two-body motions. In other words, these orbital elements describe the (osculating) two-body orbit on which the motion would continue if the perturbing forces vanished at that instant. The true motion occurs as the envelope of the osculating orbits. Naturally, in the presence of time-varying external forces such as those due to the gravitational perturbation from the second binary\footnote{We note that tidal forces may (and in several case should) be considered as well.}, the osculating elements vary continuously, with a characteristic period which is half of the orbital period of the given binary. Furthermore, if the outer orbit happens to be eccentric (as in the present case), the perturbing forces -- as well as the amplitudes of the variations of the osculating elements -- naturally increase around the periastron passage of the outer orbit. As a result, the value of a given osculating element depends strongly on the inner and outer orbital phase of the sampling. This is demonstrated in Fig.~\ref{fig:periods} by the green curve, which connects the anomalistic periods that were sampled during all the consecutive inner periastron passages of the given binary and, similarly, by the magenta curve which shows the osculating anomalistic periods during each primary eclipse.\footnote{We note that the analytical forms of these curves are described and discussed by \citet{Borkovits2015}.} However, none of the periods discussed thus far can be observed. The periods that can be observed are, e.g. the time between two consecutive primary (or, secondary) eclipses or, in case of the RV observations, the repeating time of the RV curve. As both the eclipses and the RV curve are periodic according to the true longitude-like quantity $u=v+\omega$, where $v$ denotes the true anomaly and $\omega$ stands for the argument of periastron in the observational frame of reference\footnote{For the difference between the observational and dynamical argument of periastron see, e.g., \citet{Borkovits2015}}, the measurable quantity is the $2\pi$ increment of $u$, instead of $v$, i.e. some kind of averaged sidereal (or, eclipsing) period, instead of an averaged anomalistic period. Therefore, in Fig.~\ref{fig:periods} we plot the time intervals between two consecutive primary (red) and secondary (blue) eclipses. One should be aware, however, that these observable eclipsing or, sidereal periods are systematically longer than the one-sidereal-period average of the instantaneous osculating anomalistic periods (brown symbols on the figure). The reason is that in a triple (and quadruple) system the clock of the inner binary is slowed as, on average, the perturbing forces from the outer bodies act as if the central mass of the binary were reduced. 
Finally, the black curve in Fig.~\ref{fig:periods} represents the average value of the eclipsing period which is practically the average of the red and blue curves, and can be determined most easily from the ETV curves as it is simply the linear term of the ETV ephemeris (as listed in the last row of Table~\ref{tab:Fourier_ephemeris}, and in Fig.~\ref{fig:etv}, as well).

\subsection{PHOEBE analysis}

For completeness, we also analyzed the TESS data via a classical light-curve modelling approach using the Wilson-Devinney method \citep{1971ApJ...166..605W} as implemented in the {\tt PHOEBE} code \citep{2005ApJ...628..426P}. Both binaries were analysed independently, using several simplifications for the analysis. Both orbital periods were kept fixed, and the mass ratio in each binary was set to the corresponding value given in Table \ref{tbl:simlightcurve}. Likewise, the temperature of the primary in both binaries was taken from Table \ref{tbl:simlightcurve}. See Table \ref{tbl:PHOEBElightcurve} for the results of the {\tt PHOEBE} fit. These are in good agreement with those derived with the more sophisticated approach using the {\tt Lightcurvefactory} code. Note, however, that the uncertainties associated with these {\tt PHOEBE} results are statistical only, and have often  been found to be underestimated.

\begin{table*}
\centering
\caption{Calculated parameters from \tt PHOEBE.}
\begin{tabular}{lcccc}
\hline
               & \multicolumn{4}{c}{Subsystem}  \\
               & \multicolumn{2}{c}{pair A}            & \multicolumn{2}{c}{pair B}             \\
  \hline
$i$ [deg]      & \multicolumn{2}{c}{$87.74 \pm 0.08$}  & \multicolumn{2}{c}{$87.40 \pm 0.17$}   \\
$e$            & \multicolumn{2}{c}{$0.076 \pm 0.003$} & \multicolumn{2}{c}{$0.027 \pm 0.002$}  \\  
$\omega$ [deg] & \multicolumn{2}{c}{$158.1 \pm 1.9$}   & \multicolumn{2}{c}{$187.9 \pm 5.4$}    \\
$q~[=m_\mathrm{sec}/m_\mathrm{pri}]$ & \multicolumn{2}{c}{$0.959~(\rm fixed)$} & \multicolumn{2}{c}{$0.964~(\rm fixed)$} \\
\hline
 \multicolumn{5}{c}{Derived Physical Quantities} \\
                             &       Aa            &       Ab         &        Ba          &      Bb          \\
  \hline 
 relative flux $L_{TESS}$[\%]& $29.7 \pm 1.2$      & $21.8 \pm 2.0$   & $28.3 \pm 1.5$     & $20.2 \pm 1.8$   \\
 $T_\mathrm{eff}$ [K]        & $6434~(\rm fixed) $ & $6305 \pm 26$    & $6303~(\rm fixed)$ & $6167 \pm 30$    \\
 $R$ [R$_\odot$]             & $1.281 \pm 0.008$   & $1.106 \pm 0.013$& $1.229 \pm 0.012$  &$1.061 \pm 0.017$ \\
 $M_\mathrm{bol}$            & $3.73 \pm 0.05$     & $4.14 \pm 0.07$  & $3.91 \pm 0.06$    & $4.32 \pm 0.09$  \\
 $\log g$ [dex]              & $4.300 \pm 0.013$   & $4.409 \pm 0.015$& $4.317 \pm 0.016$  & $4.430 \pm 0.017$ \\
\hline
\end{tabular}
\label{tbl:PHOEBElightcurve}
\end{table*}

\section{Discussion}
\label{sec:discussion}


\begin{figure*}
\centering
\includegraphics[width=0.49\linewidth]{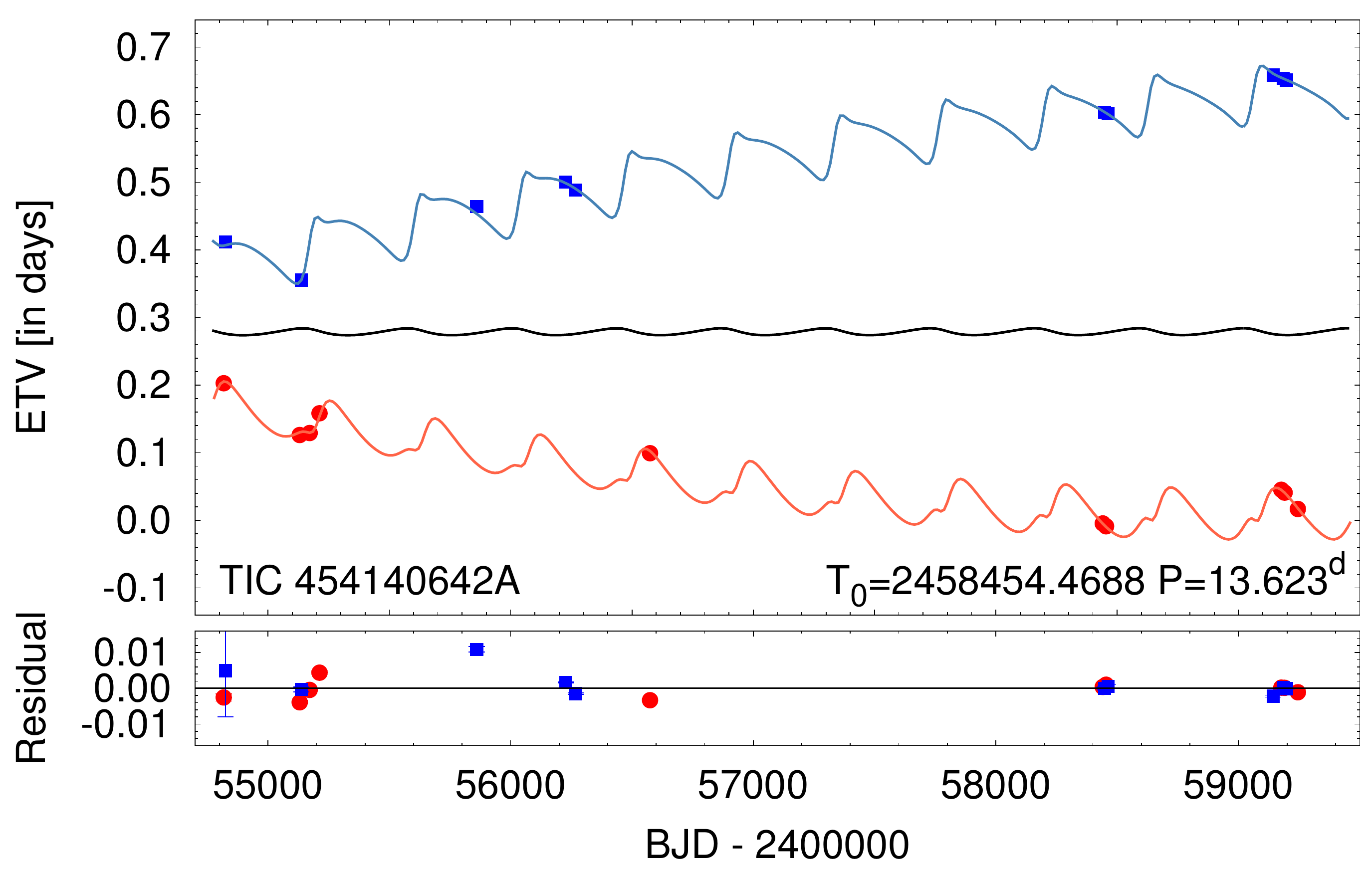} 
\includegraphics[width=0.49\linewidth]{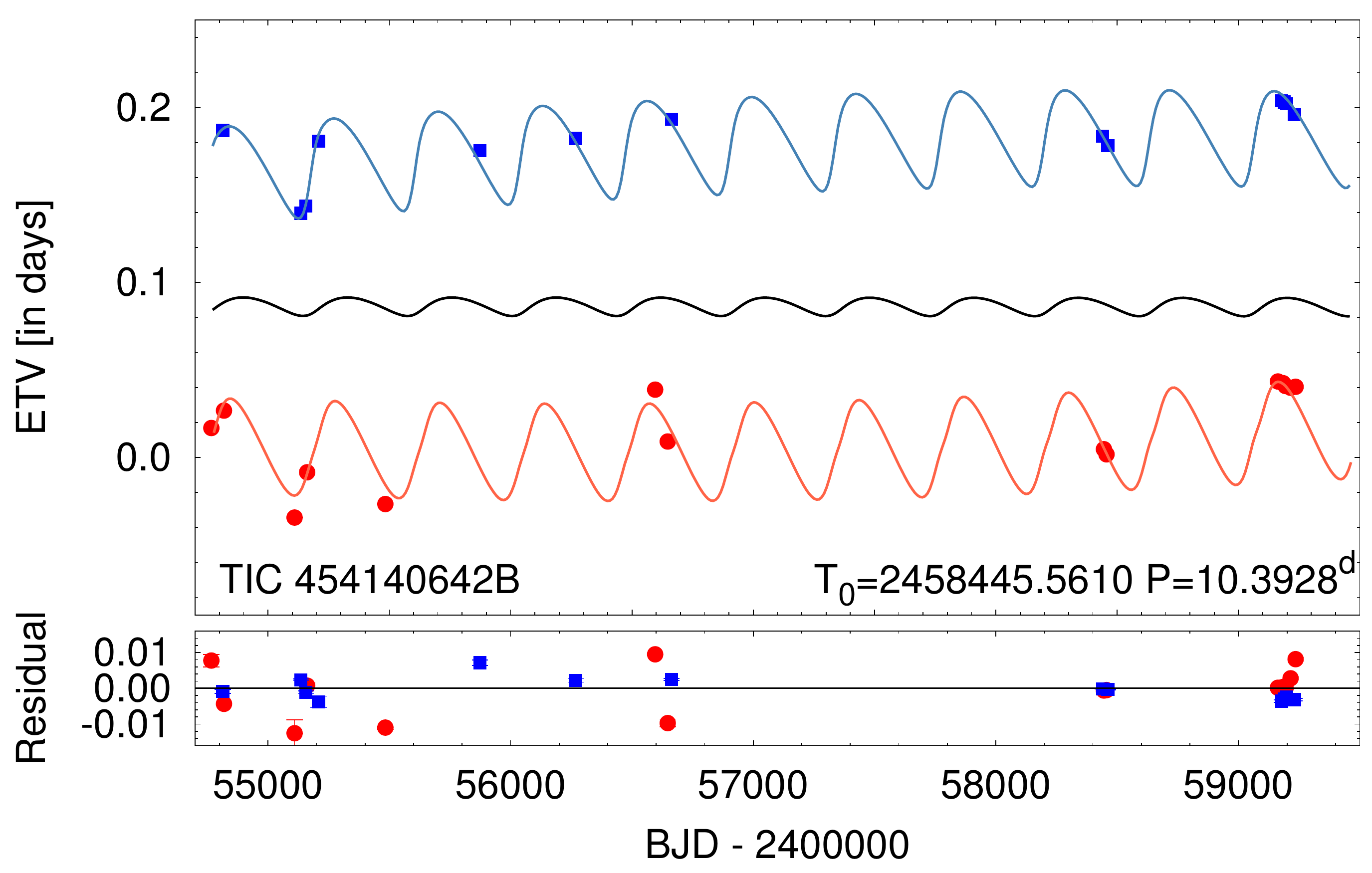}
\caption{Eclipse timing variations of binaries A and B (left and right, respectively). Red circles and blue squares represent the observed primary and secondary eclipse times, respectively, while the corresponding solid lines represent the lowest $\chi^2$ spectro-photodynamical model solution. The black lines in the middle of both upper panels represent the pure R\o mer delay (i.e., light-travel time) contribution to the given ETV. Residuals are also plotted in the lower panels.}
 \label{fig:etv} 
\end{figure*} 

\begin{figure*}
\centering
\includegraphics[width=0.95\linewidth]{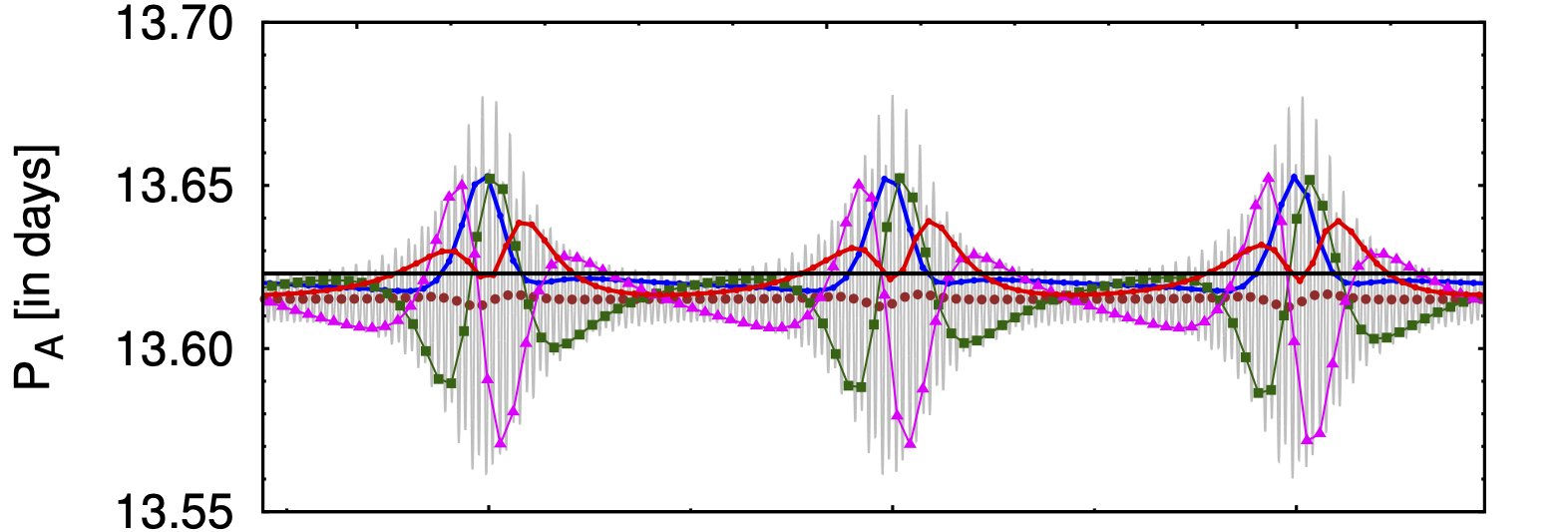} 
\includegraphics[width=0.95\linewidth]{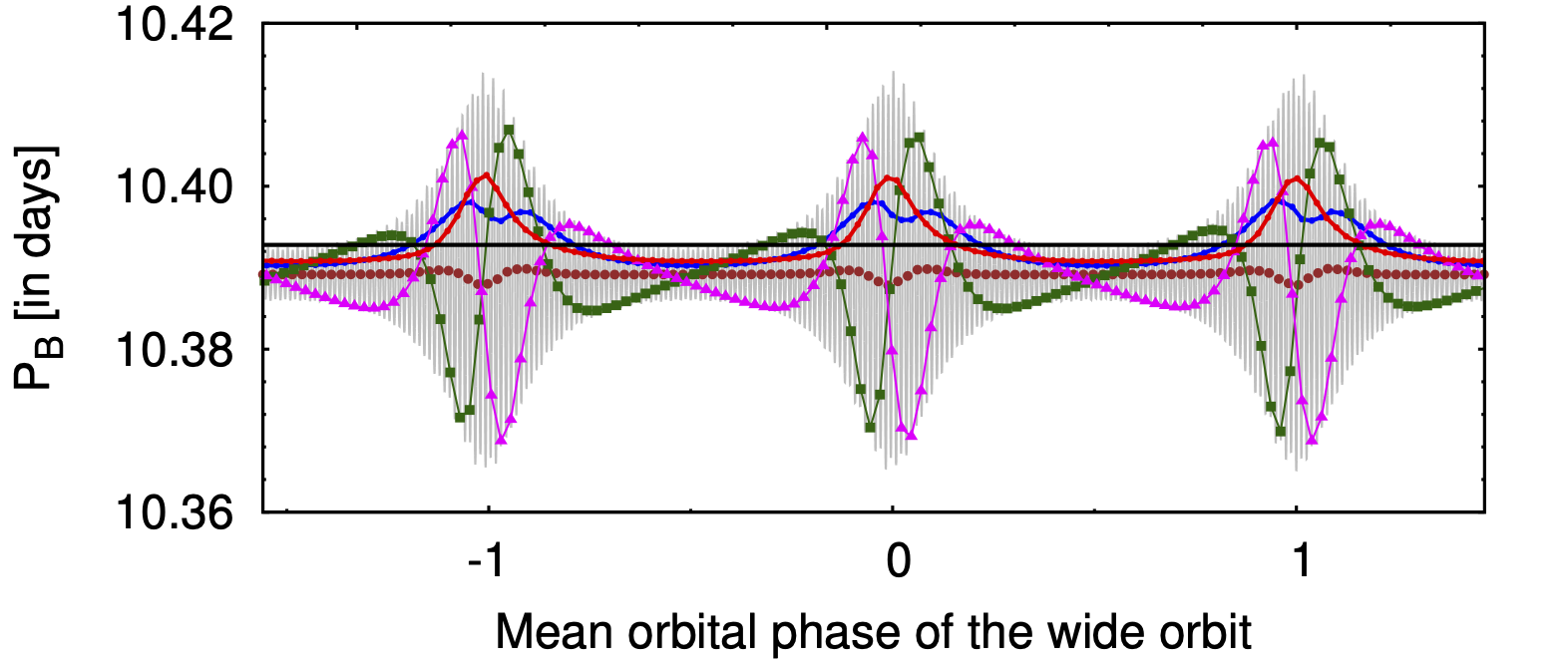}
\caption{Illustration of the various periods discussed in the text for binaries A and B (upper and lower panels, respectively). The grey curves represent the (instantaneous), osculating anomalistic periods (as derived from the osculating true anomalies via Kepler's third law). The brown circles represent the corresponding averages for one sidereal period of the binary (i.e., the interval during which the true longitude changes by $2\pi$). The dark-green squares represent the osculating anomalistic period at the time of each periastron passage of the corresponding binary, while magenta triangles denote the osculating anomalistic period at the time of the primary eclipses. The red/blue curves show the elapsed time between two primary/secondary eclipses, and can be considered to be the true, cycle-by-cycle sidereal (eclipsing) periods between two consecutive primary/secondary eclipses. Finally, the horizontal black lines represent the ``average'' sidereal (eclipsing) periods -- i.e., the periods which can be obtained from the corresponding linear term in the long-term ETV fitting; these are the periods listed in Fig.~\ref{fig:etv}.}
 \label{fig:periods} 
\end{figure*} 

\begin{figure}
\centering
\includegraphics[width=\linewidth]{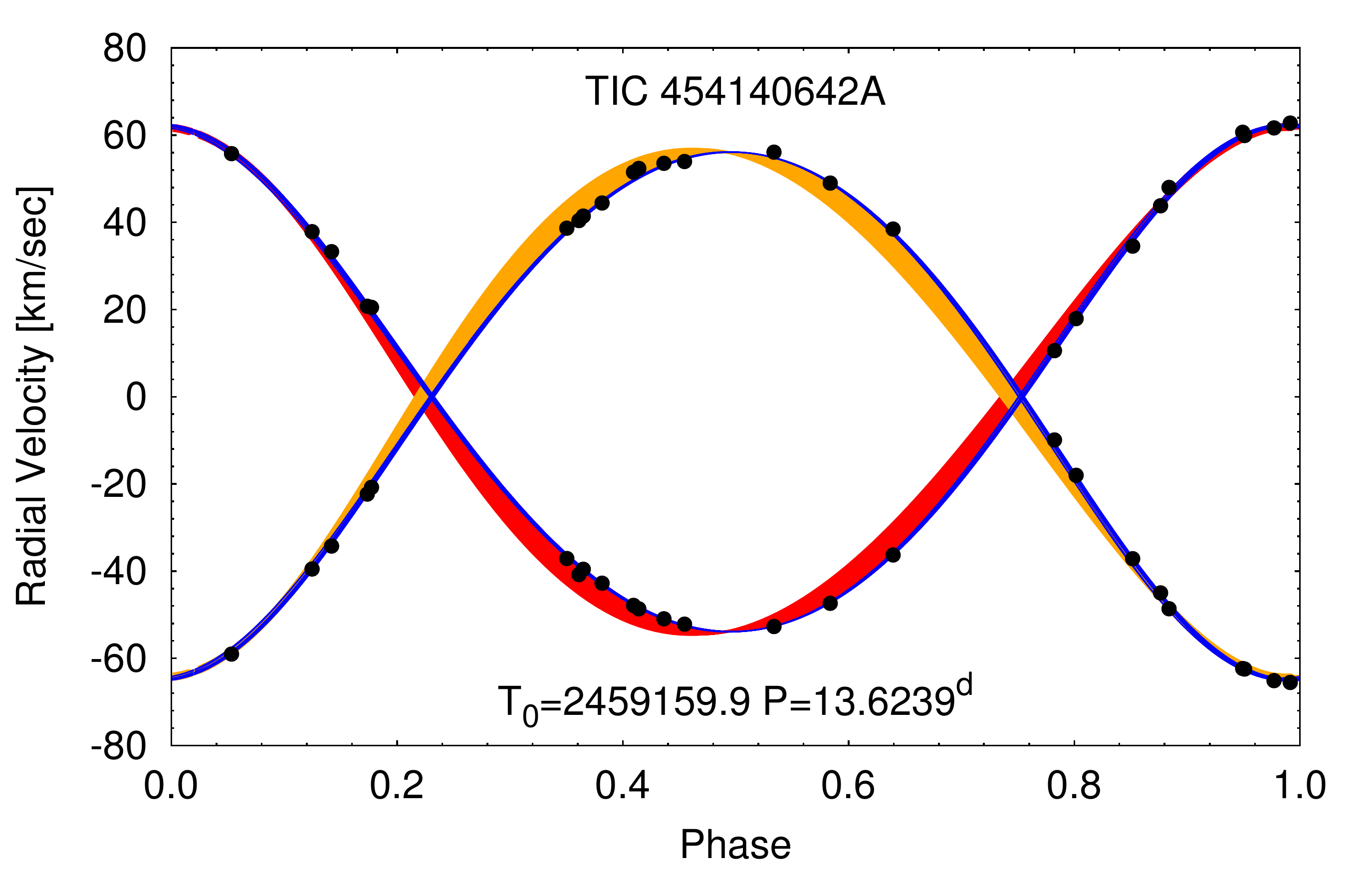} 
\caption{The folded RV curve of binary A, after the removal of the contribution of the outer orbit. Red and orange lines represent the fold of primary and secondary RV curves for the forthcoming 9.5 years. The consequence of the rapid, dynamically forced apsidal motion is clearly manifested in the varying shape of the curves. Black points represent the measured RVs (see in Table~\ref{tbl:RVs}), while the blue lines represent the folded model curves during the interval of the observations. As one can see, the half-year duration of the available observations is insufficient to detect the apsidal motion, but it will be certainly detectable within a few years.}
 \label{fig:RVAfold} 
\end{figure}

\begin{figure*}
\centering
\includegraphics[width=0.7\linewidth]{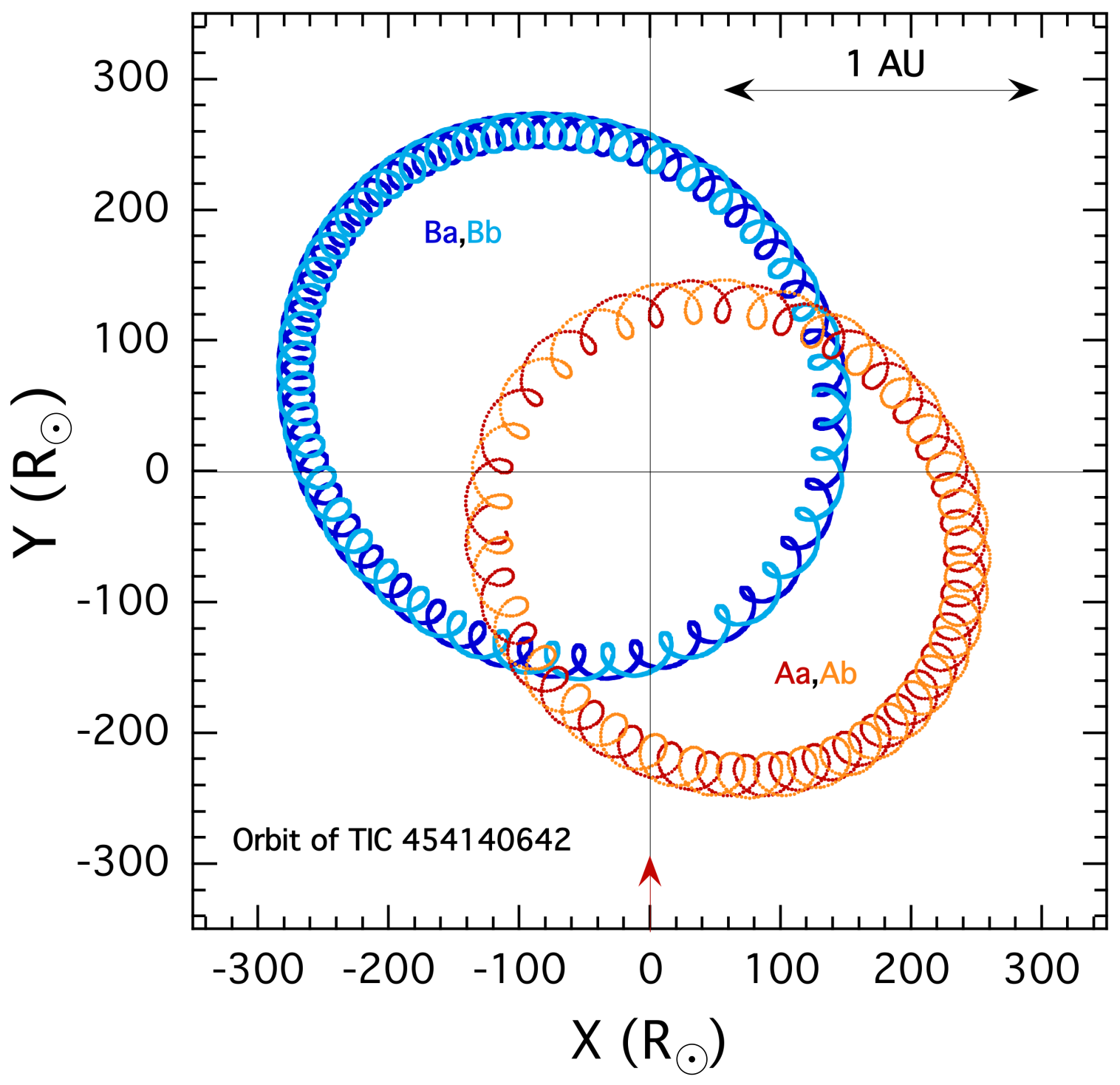} 
\caption{To-scale plot of the orbital motions in the quadruple TIC 454140642. The observer is in the $x-y$ plane and viewing the system from a distant position along the negative $y$ axis (i.e., the red arrow is the view direction).}
 \label{fig:orbit} 
\end{figure*}

\subsection{Origin}

This hierarchical system has several distinctive features pointing to its origin, namely  (i)  nearly equal masses of all four stars, (ii) a compact outer orbit, and (iii) a near-perfect coplanarity. It is well known that close binaries cannot form directly with their present-day small separations owing to the so-called 'opacity limit' to fragmentation on the order of 10 au \citep[e.g.][]{Larson1972}. Accretion of gas on to a nascent binary shrinks its orbit and at the same time increases the mass ratio because the secondary star accretes more; this mechanism can explain statistics of the close-binary population, including inner subsystems in hierarchies \citep{TokMoe2020}. Preferential growth of the secondary component produces a subset of binaries with almost equal masses (twins) which is more prominent at short periods. 
A compact quadruple such as \ticstar\ can be explained by the accretion-induced migration of both inner and outer subsystems, assuming that an initially wide and low-mass seed quadruple gained most of its mass by accretion. The seed quadruple could be produced, for example, through an encounter of two protostars A and B, each surrounded by a gas envelope. An encounter leads to a burst of accretion on to each star that produced the inner subsystems by disk fragmentation. However, other formation channels are also feasible. The important factor is a substantial mass growth of this progenitor quadruple by accretion that eventually created four stars with similar masses and shortened both inner and outer orbits. Dissipative interaction of the orbits with gas also led to the coplanarity of the entire system.  

This formation scenario of compact 2+2 quadruples makes a strong prediction of near-equal masses and orbit alignment. Indeed, other known compact quadruples share these properties with TIC 45414062. The closest such system, VW LMi (11029+3025, HD 95660), has an outer period of 355 days and inner periods of 0.478 and 7.93 days \citep{Pribulla2020}. The masses of the stars in the closest eclipsing pair are not equal, but this could be a result of the later mass transfer; the other, 7.9-day subsystem is a twin with $q=0.98$, and the outer mass ratio is 0.93; all orbits are coplanar. Another compact and doubly-eclipsing quadruple V994 Her (= HD 170314) has an outer period of 2.9 yr and inner periods of 2.1 and 1.4 days \citep{Zasche2016}. The inner mass ratios are 0.86 and 0.95, and the outer mass ratio is 0.67. There is another, more distant star in this system. Finally, the doubly-eclipsing quadruple OGLE BLG-ECL-145467, found by \citet{Zasche2019}, has an outer period of 4.2 yr, with inner periods of 3.3 and 4.9 days, and the masses of all four stars are comparable. 

\subsection{Gaia EDR3 Astrometry}

The Gaia EDR3 astrometric excess noise for TIC 454140642 is 0.144 mas at a significance of $\sigma = 24.476$; the number of good observations is 257 (Gaia EDR3, Gaia Collaboration 2020). The physical displacement corresponding to the excess noise is $\Delta a_{\rm measured} = 0.052$ au (Eqn.~3, Belokurov et al.\ 2020). Taking into account the mass and luminosity ratios between binary A and B (0.94 and 0.8, respectively), as well as the eccentricity and argument of periastron of the AB system, the expected astrometric wobble is $\Delta a_{\rm expected} = 0.053$ au (Eqn.~5, Belokurov et al.\ 2020) -- practically identical to $\Delta a_{\rm measured}$. In addition, using the methodology of Stassun \& Torres (2021) and the measured renormalized unit weight error (RUWE) of 1.299, the photocenter semi-major axis is 0.27 mas which, at the measured parallax of  2.786 mas, corresponds to $\approx0.1$ au -- comparable to $\Delta a_{\rm measured}$. Altogether, this indicates that Gaia has seen the astrometric wobble of the AB system.

\subsection{Long-term Dynamical Stability and Orbital Evolution}
\label{sect:orbitaldynamics}

The orbital elements listed in Table \ref{tbl:simlightcurve} represent a snapshot of the system at the reference epoch. Due to the strong dynamical interactions between the two binary stars, these orbital elements, in accordance with the perturbation theory of hierarchical stellar systems \citep[see, e.~g.,][]{Harrington1968} vary over time with at least three characteristic timescales: (i) the inner periods, (ii) the outer period and, (iii) the ratio of $P_\mathrm{out}^2/P_\mathrm{in}$ (see, e.g., Figs.~\ref{fig:etv}, \ref{fig:periods}). The variations on the latter two timescales can nicely be confirmed on the ETV curves (Fig.~\ref{fig:etv}), where, besides the strictly $P_\mathrm{out}$-period cyclic variations (not to be confused with the much smaller amplitude R\o mer-delay, shown also in Fig.~\ref{fig:etv}), the dynamically forced apsidal motion is also manifested.

To quantify the properties of the apsidal motions for all of the three orbits  we calculated theoretical apsidal motion rates, $\Delta\omega$ (i.e. the variation of the argument of periastron, $\omega$, during one revolution of the given binary). For this approximate calculation, we used hierarchical three-body models.  So, for example, to calculate the dynamical apsidal motion rate of binary A, we replaced the other binary (B) by a single outer body having mass $m_\mathrm{B}$ and, vice versa to compute the dynamical apsidal motion of binary B.  Besides the dynamically forced apsidal motion, we also calculated the general relativistic and tidal contributions.  For the two inner orbits we followed strictly the formulae and method described in \citet{Borkovits2015}, Appendix C.  For the outer orbit we considered the tidal contribution to be negligible and, furthermore, in the case of the dynamical term we simply calculated the algebraic sum of the contributions from the two separate `triple systems' (formed by stars Aa,Ab,B and Ba,Bb,A). We list these apsidal advance rates in the `apsidal motion parameters' part of Table~\ref{tbl:simlightcurve}.

As one can see, the dynamical apsidal advance rate ($\Delta\omega_{3b}$; where `3b' stands for `third body') is higher by 4-6 orders of magnitude than the relativistic and tidal rates ($\Delta\omega_{GR}$, $\Delta\omega_\mathrm{tide}$, respectively) for all the three orbits, therefore, these latter contributions are certainly negligible. Assuming that these apsidal motion rates remain nearly constant (which is a plausible assumption for such a nearly coplanar system) we estimated approximate theoretical apsidal motion periods for the three orbits. These periods were found to be $U_\mathrm{theo}=87\pm1$, $109\pm1$, and $480\pm1$ years for binary A, B and AB, respectively.

In order to check the plausibility of our assumptions, we determined the `true' apsidal motion periods of the three orbits. For this we numerically integrated the best-fit four-body model for $\sim3\,000$~years, and then determined the average apsidal motion periods for all three orbits. We obtained $U_\mathrm{num}=70$, $97$, and $472$~years, respectively. Therefore, we can conclude that our simple, theoretical three-body model slightly overestimated the true periods (i.e., in other words, underestimated the amplitude of the perturbations a bit); however, it resulted in a correct estimation of their magnitude.

Finally we note that the rapid apsidal motion of binary A, in addition to the readily visible divergence of the primary and secondary ETV curves, will also be detectable within a few years through the variation of the shape of the RV curve of binary A.  This is illustrated visually in Fig~\ref{fig:RVAfold}. (Note, despite the similarly rapid apsidal advance of binary B, due to its almost circular orbit, the same effect will remain below the uncertainties of the currently available RV data.)

We investigated also the short-term dynamical stability of the system. To confirm this we integrated the 4-body equations of motion for 5 million days (corresponding to $\approx10,000$ outer periods), using the parameters from Table \ref{tbl:simlightcurve}. We used the high-order \texttt{IAS15} integrator with adaptive timesteps \citep{Rein15}, part of the \texttt{REBOUND} N-body package \citep{Rein12}. We provide a \texttt{REBOUND} jupyter notebook demonstrating the initialization and orbital element computation at \url{https://github.com/vbkostov/TIC_454140642}\footnote{The orbital elements for binaries A and B represent the corresponding osculating two-body orbits, while the elements for A-B represent the orbit of the centers of mass of binaries A and B around one another.}

\begin{figure*}
    \centering
    \includegraphics[width=0.95\linewidth]{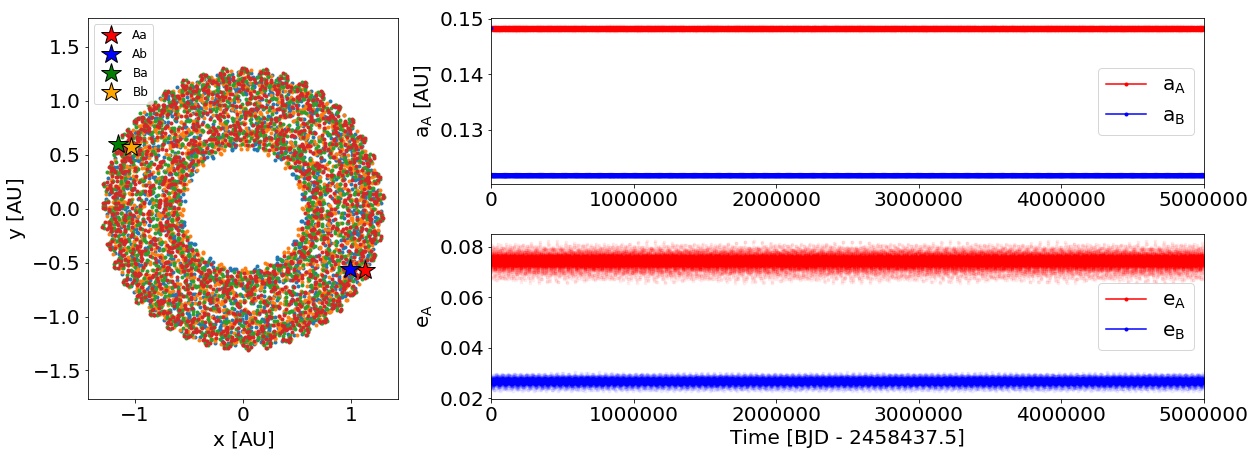}
    \caption{Left: The quadruple system as seen from above, highlighting the positions of all four stars (shown every 100 days) over the course of 200,000 days. The two binaries are separated by about an 1 au at closest approach. Upper right: Evolution of the semi-major axis of binary A and B (shown every 100 days) over the course of 5 million days (${\approx10,000}$ outer periods), showing no indications of dynamical instability for the duration of the integration. Lower right: same as upper right but for the eccentricity of binary A and B.}
    \label{fig:5milion_days_aA}
\end{figure*}

As shown in Figure \ref{fig:5milion_days_aA}, there are no indications of chaotic motion and the system remains stable for the duration of said integration. The semi-major axis of binary A ($a_A$) oscillates between 31.8 $R_\odot$ and 32 $R_\odot$ and the eccentricity ($e_A$) varies between 0.066 and 0.082; the semi-major axis of binary B ($a_B$) oscillates between 26 $R_\odot$ and 26.2 $R_\odot$, and the eccentricity ($e_B$) varies between 0.023 and 0.030. The semi-major axis of binary AB ($a_{AB}$) oscillates between 400.6 $R_\odot$ and 402.5 $R_\odot$, and the eccentricity ($e_{AB}$) varies between 0.322 and 0.328. The orbital inclinations vary between $87.48^\circ$ and $87.85^\circ$ (binary A), $87.52^\circ$ and $87.81^\circ$ (binary B), and $87.64^\circ$ and $87.69^\circ$ (binary AB). 

\begin{figure*}
    \centering
    \includegraphics[width=0.95\linewidth]{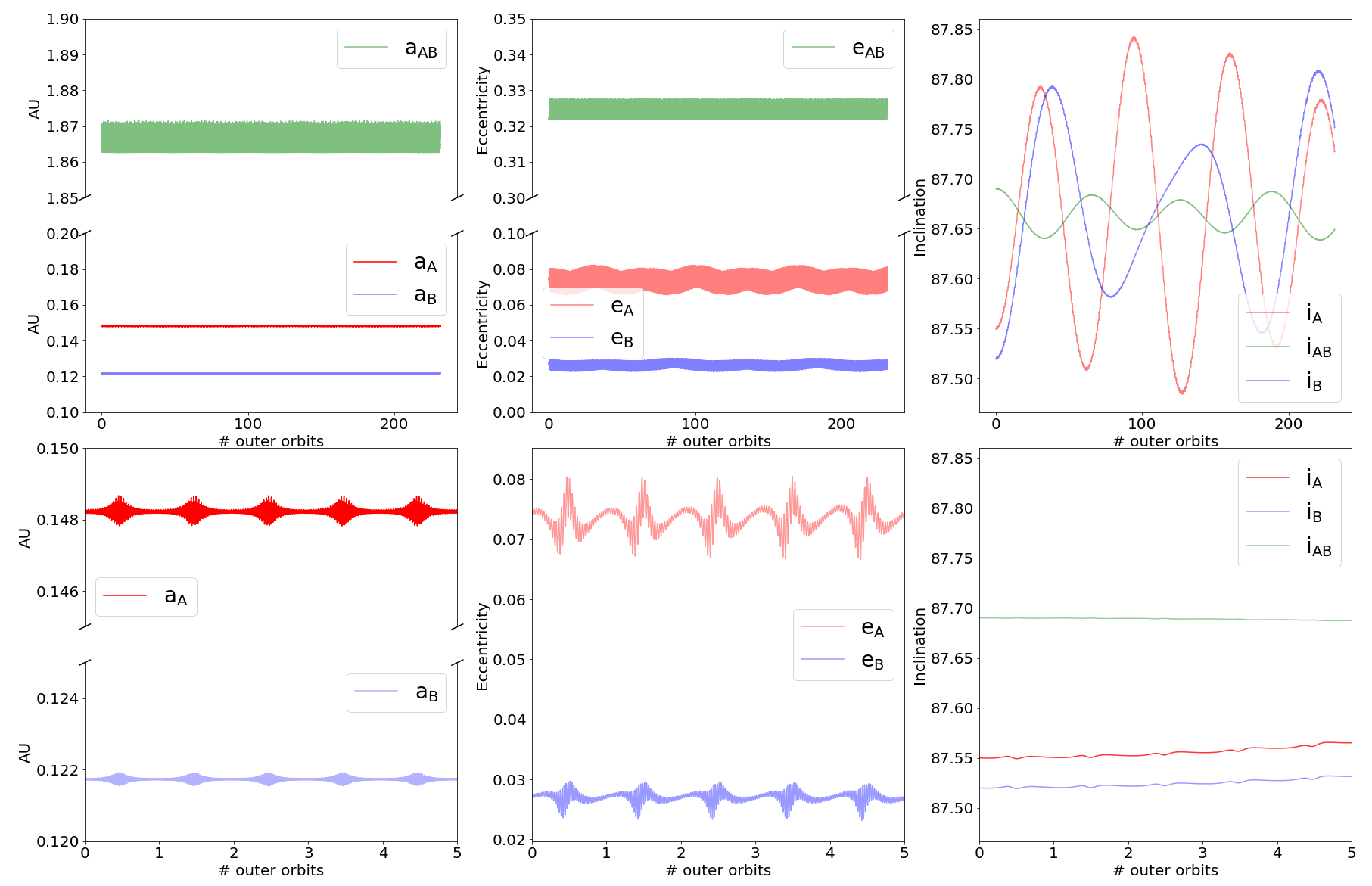}
    \caption{Upper panels: Evolution of the semimajor axis (left), eccentricity (middle), and inclination (right) of the two binaries and the quadruple over the course of 100,000 days (${\approx230}$ outer periods). Lower panels: Same as above  but zoomed-in to show the details around the 435-day outer orbit. We note that the periodic extrema in the semimajor axes and eccentricities correspond to the periastron passages of the outer orbit.}
    \label{fig:rebound_1e5_days}
\end{figure*}

\begin{figure*}
    \centering
    \includegraphics[width=0.95\linewidth]{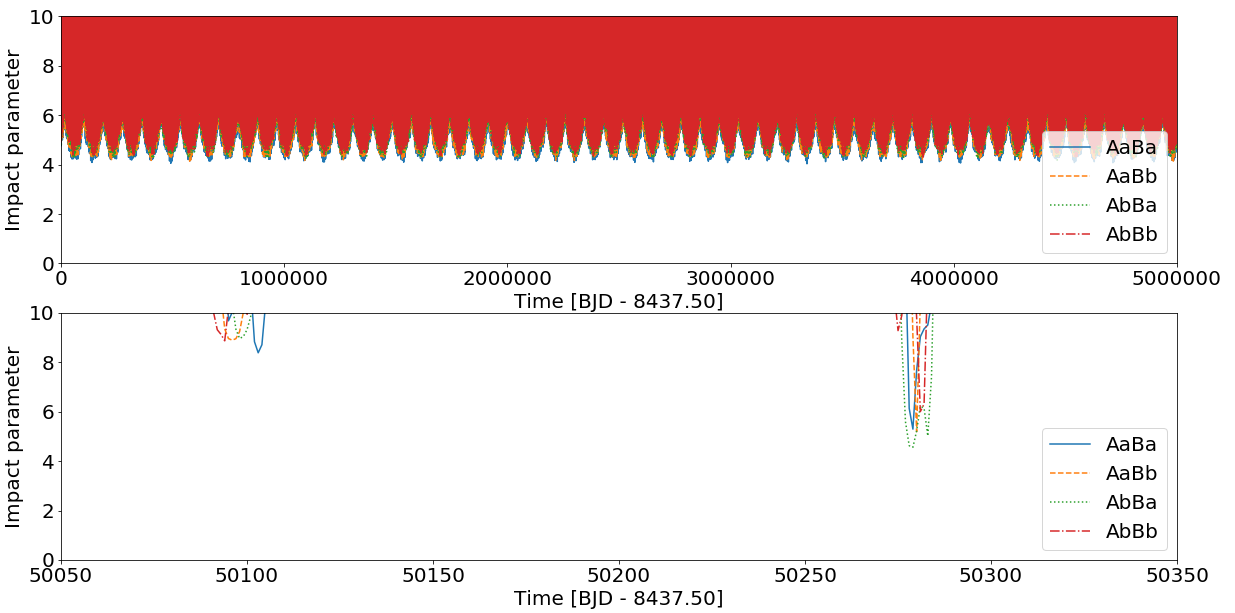}
    \caption{Upper panel: Evolution of the impact parameters between each star in each binary with respect to both stars of the other binary (e.g.~AaBa for the impact parameter between Aa and Ba, AaBb for the impact parameter between Aa and Bb, etc.) over the course of 5 millions days days. Lower panel: same as above but zoomed-in on a 300-days interval to highlight two conjunctions. }
    \label{fig:impact_param}
\end{figure*}

Despite the near edge-on orientation of the system, there are no mutual eclipses or occultations between the constituents stars of binary A and binary B. This is highlighted in Figure \ref{fig:impact_param} where we show the evolution of the respective impact parameters between the four stars. While there are periodic, precession-induced variations in the corresponding minimum impact parameters, the latter never reach below $\approx4$. 

\section{Summary}
\label{sec:summary}

We have presented the discovery of the quadruple-lined, doubly eclipsing quadruple system TIC 454140642. The target was observed by TESS in Sector 5 (at 30-min cadence) and again in Sector 32 (in 10-min cadence), and exhibited four sets of eclipses associated with two binary stars (A and B). Additional spectroscopic and photometric measurements reveal divergent radial velocities, eclipse-time variations and apsidal motion for both binaries, confirming that they form a gravitationally-bound quadruple system with a 2+2 hierarchy. The two binary stars have orbital periods of $P_A = 13.62$ days and $P_B = 10.39$ days, respectively, and slightly eccentric orbits ($e_A = 0.07$, $e_B = 0.03$). The four stars are very similar in terms of mass (individual masses between $1.105 M_\odot$ and $1.195 M_\odot$), size (radii between $1.09 R_\odot$ and $1.26 R_\odot$), and effective temperature (between 6188 K and 6434 K). The quadruple has an orbital period of $P_{AB} = 432$ days and an eccentricity of $e_{AB} = 0.32$. The entire system is essentially coplanar -- the orbits of the two binaries and the wide orbit are aligned to within a fraction of a degree. \ticstar\ is the newest addition to the small family of confirmed, well-characterized, multiply-eclipsing quadruple systems. 

\acknowledgments
This paper includes data collected by the \emph{TESS} mission, which are publicly available from the Mikulski Archive for Space Telescopes (MAST). Funding for the \emph{TESS} mission is provided by NASA's Science Mission directorate. This work makes use of observations from the LCOGT network.

Resources supporting this work were provided by the NASA High-End Computing (HEC) Program through the NASA Center for Climate Simulation (NCCS) at Goddard Space Flight Center.  Personnel directly supporting this effort were Mark L. Carroll, Laura E. Carriere, Ellen M. Salmon, Nicko D. Acks, Matthew J. Stroud, Bruce E. Pfaff, Lyn E. Gerner, Timothy M. Burch, and Savannah L. Strong.

This research has made use of the Exoplanet Follow-up Observation Program website, which is operated by the California Institute of Technology, under contract with the National Aeronautics and Space Administration under the Exoplanet Exploration Program. 

This research is based in part on observations made with the Galaxy Evolution Explorer, obtained from the MAST data archive at the Space Telescope Science Institute, which is operated by the Association of Universities for Research in Astronomy, Inc., under NASA contract NAS5-26555.

We would also like to thank the Pierre Auger Collaboration for the use of its facilities. The operation of the robotic telescope FRAM is supported by the grant of the Ministry of Education of the Czech Republic LM2018102. The data calibration and analysis related to the FRAM telescope is supported by the Ministry of Education of the Czech Republic MSMT-CR LTT18004, MSMT/EU funds CZ.02.1.01/0.0/0.0/16$\_$013/0001402 and CZ.02.1.01/0.0/0.0/18$\_$046/0016010.

This work is supported by MEYS (Czech Republic) under the projects MEYS LM2018505, LTT17006 and EU/MEYS CZ.02.1.01/0.0/0.0/16$\_$013/0001403 and CZ.02.1.01/0.0/0.0/18$\_$046/0016007.

TB acknowledges the financial support of the Hungarian National Research, Development and Innovation Office -- NKFIH Grant KH-130372.

This work has made use of data from the European Space Agency (ESA) mission {\it Gaia} (\url{https://www.cosmos.esa.int/gaia}), processed by the {\it Gaia} Data Processing and Analysis Consortium (DPAC, \url{https://www.cosmos.esa.int/web/gaia/dpac/consortium}). Funding for the DPAC has been provided by national institutions, in particular the institutions participating in the {\it Gaia} Multilateral Agreement.

Resources supporting this work were provided by the NASA High-End Computing (HEC) Program through the NASA Advanced Supercomputing (NAS) Division at Ames Research Center for the production of the SPOC data products.

This work makes use of observations from the LCOGT network.

\facilities{
\emph{Gaia},
MAST,
TESS,
WASP,
ASAS-SN,
NCCS,
FRAM-Auger,
FRAM-CTA,
PEST,
TRES,
SOAR,
LCOGT}

\software{
{\tt Astrocut} \citep{astrocut},
{\tt AstroImageJ} \citep{Collins:2017},
{\tt Astropy} \citep{astropy2013,astropy2018}, 
{\tt Eleanor} \citep{eleanor},
{\tt IPython} \citep{ipython},
{\tt Keras} \citep{keras},
{\tt Keras-vis} \citep{kerasvis}
{\tt Lightcurvefactory} \citep{Borkovits2013,Rappaport2017,Borkovits2018},
{\tt Lightkurve} \citep{lightkurve},
{\tt Matplotlib} \citep{matplotlib},
{\tt Mpi4py} \citep{mpi4py2008},
{\tt NumPy} \citep{numpy}, 
{\tt Pandas} \citep{pandas},
{\tt PHOEBE} \citep{2011ascl.soft06002P},
{\tt Scikit-learn} \citep{scikit-learn},
{\tt SciPy} \citep{scipy},
{\tt Tapir} \citep{Jensen:2013},
{\tt Tensorflow} \citep{tensorflow},
{\tt Tess-point} \citep{tess-point}
}

\bibliography{refs}{}
\bibliographystyle{aasjournal}



\end{document}